\documentclass[aps,prb, showpacs,superscriptaddress,longbibliography]{revtex4-2}  %

\usepackage{graphicx}  
\usepackage{dcolumn}   
\usepackage{bm}        
\usepackage{amssymb}   
\usepackage{amsmath}
\usepackage{gensymb}
\usepackage{color}  
\usepackage{hyperref}
\usepackage{layouts}
\usepackage{braket}
\usepackage[dvipsnames]{xcolor}
\usepackage[normalem]{ulem}
\usepackage[separate-uncertainty=true,multi-part-units=single]{siunitx}

\setcitestyle{super,open={},close={}}

\renewcommand{\vec}[1]{\boldsymbol{\mathbf{#1}}}

\newcommand \Gone {G1}
\newcommand \Gtwo {G2}
\newcommand \Gthree {G3}
\newcommand \Gfour {G4}
\newcommand \Qtwo {QD2}
\newcommand \phid {\phi_{\rm drain}}
\newcommand \Vgone {V_{\rm G1}}
\newcommand \Vgtwo {V_{\rm G2}}
\newcommand \Vgthree {V_{\rm G3}}
\newcommand \Vgfour {V_{\rm G4}}
\newcommand \LSESone {{\rm LSES}_{\rm G1}}
\newcommand \LSEStwo {{\rm LSES}_{\rm G2}}
\newcommand \LSESthree {{\rm LSES}_{\rm G3}}

\newcommand{\Shfi}[1]{S_{{\rm G}#1}^{\rm hf}}
\DeclareMathOperator{\Slf}{\mathnormal{S}_{G\mathnormal{i}}^{lf}}
\DeclareMathOperator{\Shf}{\mathnormal{S}_{G\mathnormal{i}}^{hf}}
\DeclareMathOperator{\gt}{\mathsf{g}}
\DeclareMathOperator{\filter}{\mathnormal{\eta_t}}
\DeclareMathOperator{\ffilter}{\mathnormal{\tilde{\eta}_t}}
\DeclareSIUnit{\belmilliwatt}{Bm}
\DeclareSIUnit{\dBm}{\deci\belmilliwatt}

\hypersetup{
    colorlinks,
    linkcolor={blue!50!black},
    citecolor={blue!50!black},
    urlcolor={blue!80!black}
}

\DeclareGraphicsExtensions{.pdf,.png}

\begin{document}

\title{A single hole spin with enhanced coherence in natural silicon}

\author{N. Piot}
\thanks{Contributed equally to the work.}
\author{B. Brun}
\thanks{Contributed equally to the work.}
\author{V. Schmitt}
\author{S. Zihlmann}
\affiliation{Univ. Grenoble Alpes, CEA, Grenoble INP, IRIG-Pheliqs, Grenoble, France.}
\author{V. P. Michal}
\affiliation{Univ. Grenoble Alpes, CEA, IRIG-MEM-L\_Sim, Grenoble, France.}
\author{A. Apra}
\affiliation{Univ. Grenoble Alpes, CEA, Grenoble INP, IRIG-Pheliqs, Grenoble, France.}
\author{J. C. Abadillo-Uriel}
\affiliation{Univ. Grenoble Alpes, CEA, IRIG-MEM-L\_Sim, Grenoble, France.}
\author{X. Jehl}
\affiliation{Univ. Grenoble Alpes, CEA, Grenoble INP, IRIG-Pheliqs, Grenoble, France.}
\author{B. Bertrand}
\author{H. Niebojewski}
\author{L. Hutin}
\author{M. Vinet}
\affiliation{Univ. Grenoble Alpes, CEA, LETI, Minatec Campus, Grenoble, France.}
\author{M. Urdampilleta}
\affiliation{Univ. Grenoble Alpes, CNRS, Grenoble INP, Institut Néel, Grenoble, France.}
\author{T. Meunier}
\affiliation{Univ. Grenoble Alpes, CNRS, Grenoble INP, Institut Néel, Grenoble, France.}
\author{Y.-M. Niquet}
\affiliation{Univ. Grenoble Alpes, CEA, IRIG-MEM-L\_Sim, Grenoble, France.}
\author{R. Maurand}
\email{romain.maurand@cea.fr}
\author{S. De Franceschi}
\email{silvano.defranceschi@cea.fr}
\affiliation{Univ. Grenoble Alpes, CEA, Grenoble INP, IRIG-Pheliqs, Grenoble, France.}

\date{\today}

\email{boris.brun-barriere@cea.fr, romain.maurand@cea.fr,silvano.defranceschi@gmail.com}

\begin{abstract}
\textbf{Semiconductor spin qubits based on spin-orbit states are responsive to electric field excitation allowing for practical, fast and potentially scalable qubit control. Spin-electric susceptibility, however, renders these qubits generally vulnerable to electrical noise, which limits their coherence time.  Here we report on a spin-orbit qubit consisting of a single hole electrostatically confined in a natural silicon metal-oxide-semiconductor device. By varying the magnetic field orientation, we reveal the existence of operation sweet spots where the impact of charge noise is minimized while preserving an efficient electric-dipole spin control.  We correspondingly observe an extension of the Hahn-echo coherence time up to $88$\,$\mu$s, exceeding by an order of magnitude the best values reported for hole-spin qubits, and approaching the state-of-the-art for electron spin qubits with synthetic spin-orbit coupling in isotopically-purified silicon. This finding largely enhances the prospects of silicon-based hole spin qubits for scalable quantum information processing.  
}

\end{abstract}

\maketitle


\section*{Introduction}
In the global effort to build scalable quantum processors, spin qubits in semiconductor quantum dots\cite{Loss_PRA_1998} are progressively making their mark
\cite{burkard2021}. We highlight, in particular, the achievement of single- \cite{Veldhorst-2014,Yoneda-2018} and two-qubit  
\cite{Huang-2019, Noiri_Nature_2022, Xue_Nature_2022, mills2021} 
gate fidelities well above 99\%,
the first realizations of multi-qubit arrays \cite{takeda2021,Veldhorst_2021}, and a demonstrated compatibility with industrial-grade semiconductor manufacturing technologies \cite{Maurand_Ncomms_2016,Zwerver_2022,Zalba_natcom_2021}.

Owing to their long coherence time, electron-spin qubits in silicon quantum dots have so far attracted the largest attention \cite{burkard2021}. That said, their control requires add-ons such as  metal microstrips \cite{Veldhorst-2014}, micromagnets \cite{Yoneda-2018}, or dielectric resonators \cite{Vahapoglu_SciAdv_2021}, whose large-scale integration is technically challenging
\cite{Zalba_natcom_2021}. 
Hole-spin qubits, on the other hand, can circumvent this difficulty thanks to their intrinsically large spin-orbit coupling, which enables electric-dipole spin manipulation. Over the last five years a variety of hole spin qubits have been reported in both silicon \cite{Maurand_Ncomms_2016,Camenzind_ArxiV_2021} and germanium \cite{Watzinger_Ncomms_2018, Jirovec2021, Froning_NNano_2021, Scappucci2020} quantum dots. In all these qubits, quantum operations are performed using high-frequency gate voltage excitations. 

The downside of all-electrical spin control is that the required spin-orbit coupling exposes the qubit to charge noise, leading to a reduced hole spin coherence. Recent theoretical works\cite{malkoc_2022_arxiv,Bosco_PRXQ_2021, Wang_NPJQI_2021}, however, have shown that, for properly chosen structural geometries and magnetic field orientations, careful tuning of the electrostatic confinement can bring the hole qubit to an optimal operation point where the effects of charge noise vanish to first order while enabling efficient electric-dipole spin resonance. Here, using a single hole spin confined in natural silicon we pinpoint the existence of operation sweet-spots where the longitudinal spin-electric susceptibility is minimized, resulting in a large enhancement of the spin coherence time.

\begin{figure*}[]
\includegraphics[width = 1 \textwidth]{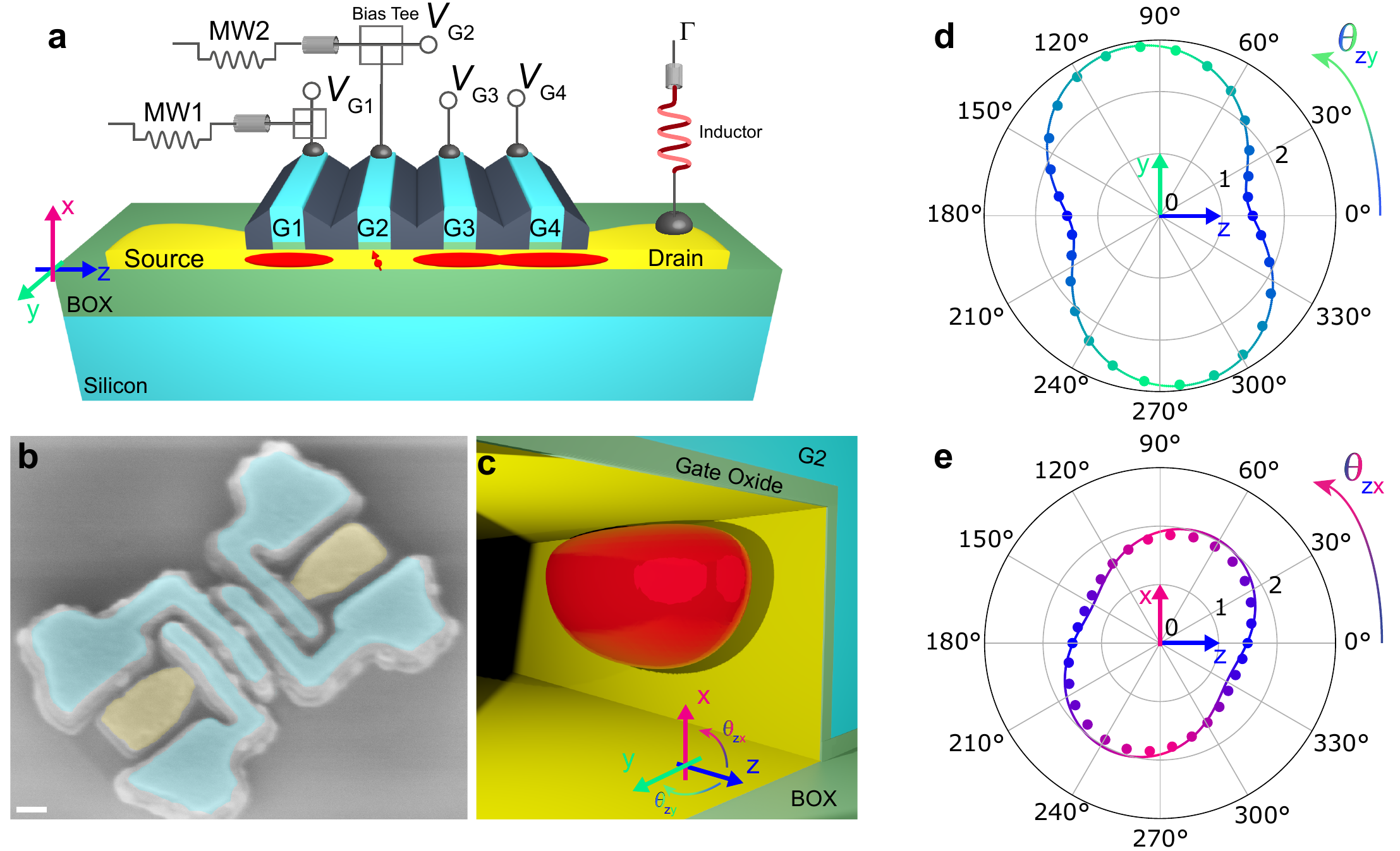}
\caption{\label{fig1} \textbf{Device, measurement scheme, and properties of the first confined hole.}
\textbf{(a)} Simplified 3-dimensional representation of a silicon (yellow)-on-insulator (green) nanowire device with four gates (light blue) labelled \Gone{}, \Gtwo{}, \Gthree{} and \Gfour{}. Gate \Gtwo{} defines a quantum dot (\Qtwo{}) hosting a single hole; \Gthree{} and \Gfour{} define a hole island used as reservoir and sensor for hole spin readout; 
\Gone{} defines a hole island
screening \Qtwo{} from dopant disorder and fluctuations in the source.
Using bias-tees, both static voltages ($\Vgone$, $\Vgtwo$) and time-dependent, high-frequency voltages (MW1, MW2) can be applied to \Gone{} and \Gtwo{}, respectively. The drain contact is connected to an off-chip, surface-mount inductor to enable rf reflectometry readout. The coordinate system used for the magnetic field is shown on the left side (in the crystal frame, $x=[001]$, $y=[1\Bar{1}0]$ and $z=[110]$). Each axis is given a different color, which is used throughout the manuscript to indicate the magnetic field orientation.  \textbf{(b)} Colorized scanning electron micrograph showing a tilted view of a device similar to the measured one. Image taken just after the etching of the spacer layers. Scale bar: $100$\,nm  \textbf{(c)} Artistic view of the calculated wave function of the first hole accumulated under \Gtwo{}. \textbf{(d)} Measured (dots) and calculated (solid line) hole $g$-factor as a function of the in-plane magnetic field angle $\theta_{zy}$ (dots). $\theta_{zy}=90\degree$ corresponds to a magnetic field applied along the $y$ axis. \textbf{(e)} Same as (d) in the $xz$ plane. $\theta_{zx}=90\degree$ corresponds to a magnetic field applied along the $x$ axis.}
\end{figure*}

\section*{Device and $g$-factors}

Our device consists of an undoped silicon nanowire with rectangular cross-section 
whose electrostatics is controlled by four gates (\Gone{} to \Gfour{}) as shown in Figs.~\ref{fig1}a,b (see Methods for details). We define a large hole island below \Gthree{} and \Gfour{} to be used simultaneously as a reservoir and as a charge sensor for a single hole trapped in a quantum dot, \Qtwo{}, under \Gtwo{}. Single-shot readout of this hole spin is performed by means of a spin-to-charge conversion technique based on the real-time detection of spin-selective tunneling to the reservoir, a widely used method often referred to as ``Elzerman readout'' \cite{Chatterjee_Natrev_2021}. Tunneling events are detected by dispersive rf-reflectometry on the charge sensor (see Methods and Supp. Info S1 for technical details). 

In our device geometry, the first holes primarily accumulate in the upper corners of the Si nanowire \cite{Voisin_NL_2014}. Figure~\ref{fig1}c displays the expected single hole wave function in \Qtwo{}, computed with a finite-differences $\vec{k}\cdot\vec{p}$ model including the six topmost valence bands \cite{Venitucci_PRB_2018} (see Supp. Info S2). At low-energy, i.e. close to the valence-band edge, the hole wave function primarily contains heavy-hole (HH) and light-hole (LH) components. 

The strong two-axes confinement readily seen in Fig.~\ref{fig1}c favors HH-LH mixing \cite{Kloeffel_PRB_2018, Michal_PRB_2021}. This mixing is expected to manifest in the anisotropy of the hole $g$-tensor, which bears information on the relative weight of HH and LH components  \cite{Zwanenburg_NL_2009,Ares_PRL_2013,Bogan2017}. To verify this, we measure the hole spin resonance frequency $f_L$ while varying the orientation of the magnetic field $\vec{B}$ in the $xz$ and $yz$ planes. The effective $g$-factor $g=hf_L/(\mu_B|\vec{B}|)$ (with $\mu_B$ the Bohr magneton and $h$ the Planck constant) is plotted in Figs.~\ref{fig1}d and 1e as a function of the magnetic field angles $\theta_{zx}$ and $\theta_{zy}$, respectively. These maps highlight the strong anisotropy of the Zeeman splitting, with a maximal $g=2.7$ close to the $y$ axis (in-plane, perpendicular to the wire) and a minimal $g=1.4$ close the $z$ axis (in-plane, along the wire). The calculated $g$-factors are also plotted in the same figures as colored solid lines. The agreement with the experimental data is remarkable. From the numerical simulation, we conclude that the measured $g$-factor anisotropy results from a strong electrical confinement against the side facet of the channel (along $y$), which prevails over the mostly structural vertical confinement (along $x$). The experimental $g$-factors and the small misalignment between the principal axes of the $g$-tensor and the device symmetry axes are best reproduced by introducing a moderate amount of charge disorder in combination with small ($\sim 0.1$\%) shear strains in the silicon channel (see Methods and Supp. Info S2). The latter likely originate from device processing and thermal contraction at the measurement temperature \cite{liles_2020}.

\begin{figure*}[]
\includegraphics[width = 1 \textwidth]{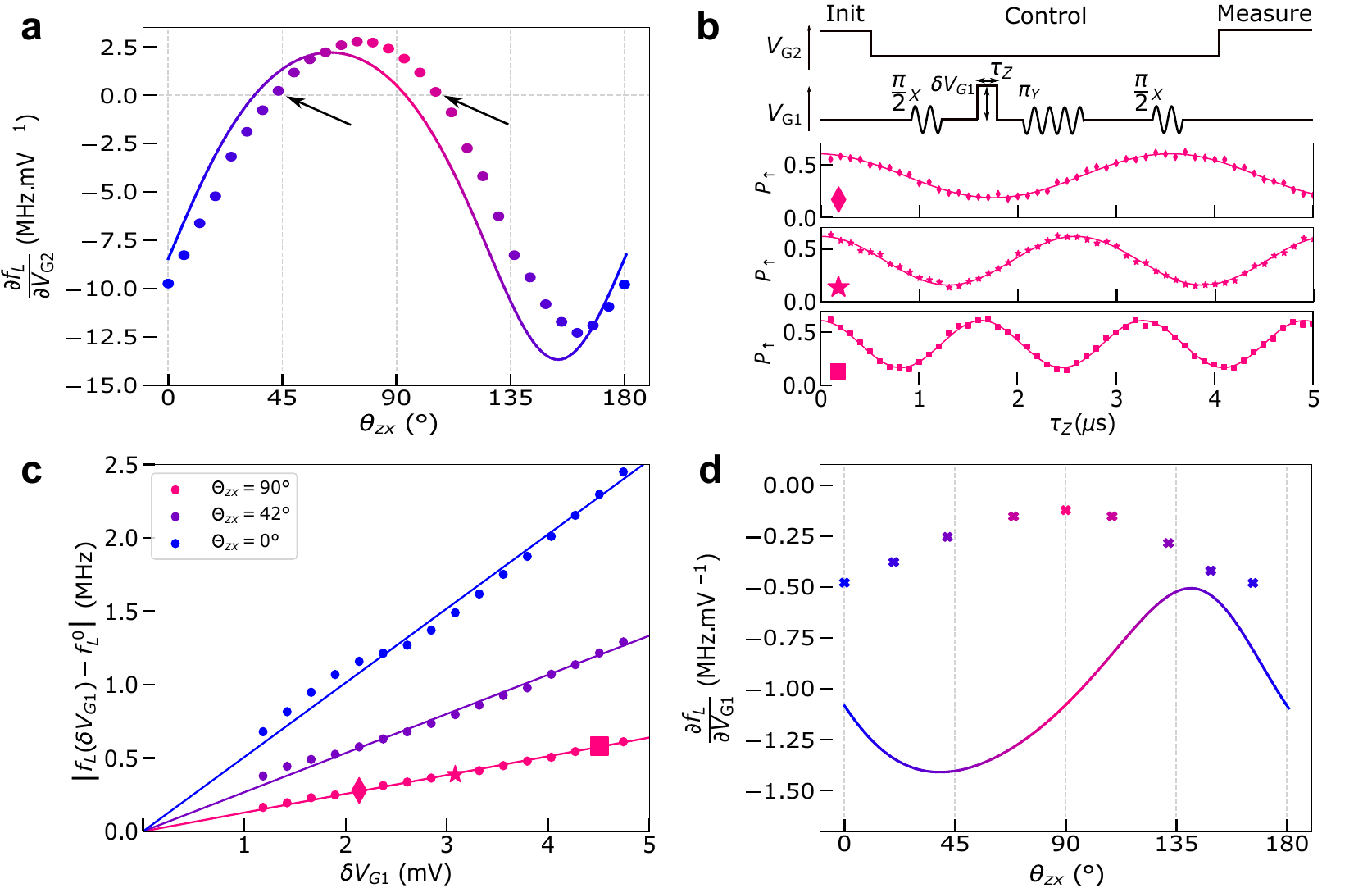}
\caption{\label{fig2}\textbf{Longitudinal spin-electric susceptibility (LSES).} \textbf{(a)} Spin-electric susceptibility with respect to $\Vgtwo$ ($\LSEStwo$) as a function of magnetic field angle $\theta_{zx}$ (symbols), at constant $f_L=19$\,GHz. The LSES vanishes at $\theta_{zx}=41\degree$ and $106\degree$, as indicated by the two arrows. The solid line corresponds to the numerically calculated $\LSEStwo$. 
\textbf{(b)} (top) Pulse sequence used to measure $\LSESone$: a voltage pulse of amplitude $\delta V_{\rm G1}$ and duration $\tau_{\rm Z}$ is applied to \Gone{} during the first free evolution time of a Hahn-echo sequence. (bottom) Spin-up fraction $P_{\uparrow}$ as a function of $\tau_{\rm Z}$ for $\delta V_{\rm G1}=2.16$\,mV (diamonds), $3.12$\,mV (stars) and $4.80$\,mV (squares), at $\theta_{zx}= 90\degree$. The oscillation frequency varies with $\delta V_{\rm G1}$. \textbf{(c)} $\delta V_{\rm G1}$ dependence of the frequency shift extracted from the Hahn-echo measurements at $\theta_{zx}=0\degree$, $42\degree$ and $90\degree$. Symbols in the latter data set correspond to the $P_{\uparrow}$ oscillations shown in (b). The solid lines are linear fits to the experimental data whose slope directly yields $|\LSESone|$. \textbf{(d)} Measured  (symbols) and calculated (solid line) $\LSESone$ as a function of $\theta_{zx}$, at constant $f_L=17$\,GHz. The negative sign of $\LSESone$ is inferred from the shift of $f_L$ under a change in $\Vgone$.}
\end{figure*}

\section*{Longitudinal spin-electric susceptibility}

Given that the $g$-factor anisotropy is intimately related to the HH/LH mixing, which is controlled by the electrostatic confinement potential, the Larmor frequency is expected to be gate-voltage dependent. As a consequence, the hole spin coherence must be generally susceptible to charge noise. We thus measure the longitudinal spin-electric susceptibility (LSES) with respect to the voltages applied to the lateral gate \Gone{} and to the accumulation gate \Gtwo{}, which we define as $\LSESone=\dfrac{\partial f_L}{\partial \Vgone}$ and $\LSEStwo=\dfrac{\partial f_L}{\partial \Vgtwo}$, respectively. In essence, $\LSESone$ and $\LSEStwo$ characterize the response of the Larmor frequency to the electric-field components parallel  ($z$) and perpendicular ($x,y$) to the channel direction, respectively. 

To probe the response to \Gtwo{}, we directly measure the spin resonance frequency $f_L$ at different $\Vgtwo$ (see Supp. Info S3 for details). The resulting $\LSEStwo$ is plotted as a function of the magnetic field angle $\theta_{zx}$ in Fig.~\ref{fig2}a. The observed angular dependence is in good agreement with the theoretical expectation. 

Noticeably, $\LSEStwo$ is positive along $x$ and negative along $z$. Indeed, when increasing $\Vgtwo$, the hole wave function extends proportionally more in the $yz$ plane than in the vertical $x$ direction, which increases $g_x$ and decreases $g_y$ and $g_z$ (see Supp. Info S2). As a result of the sign change, $\LSEStwo$ vanishes at two magnetic field orientations in the $xz$ plane (marked by arrows in Fig.~\ref{fig2}a), which are sweet-spots for electric-field fluctuations perpendicular to the silicon channel. 

To probe the response to \Gone{}, we introduce a pulse on $\Vgone$ in a Hahn-echo sequence~\cite{Yoneda-2018} as outlined in Fig.~\ref{fig2}b. This defines a phase gate, controlled by the amplitude $\delta V_{\rm G1}$ and duration $\tau_{\rm Z}$ of the pulse. Figure~\ref{fig2}b displays the coherent oscillations recorded as a function of $\tau_{\rm Z}$ for three different pulse amplitudes. The frequency of these oscillations is expected to increase linearly with $\delta V_{\rm G1}$, with a slope $\LSESone=\frac{\partial f_L}{\partial \Vgone}$. This is shown in Fig.~\ref{fig2}c for different magnetic field orientations. $\LSESone$, plotted in Fig.~\ref{fig2}d as a function of $\theta_{zx}$, ranges from $-0.5$\,MHz/mV to $-0.1$\,MHz/mV.  Its magnitude is much smaller than $\LSEStwo$ because \Gone{} is farther from \Qtwo{} than \Gtwo{} and its field effect is partly screened by the hole gas beneath. The numerically calculated $\LSESone$ (solid line) reproduces reasonably well the order of magnitude but not the angular dependence of the measured $\LSESone$. This discrepancy may be due to inaccuracies in the description of the hole gases near QD2 as well as to unaccounted charge disorder and strains  (see discussion in Supp. Info S2). We also notice that $\LSESone$ never vanishes and that the minimum of $|\LSESone|$ happens to be almost at the same $\theta_{zx}$ as a zero of $\LSEStwo$. 

\begin{figure*}[ht!]
\includegraphics[width = 1 \textwidth]{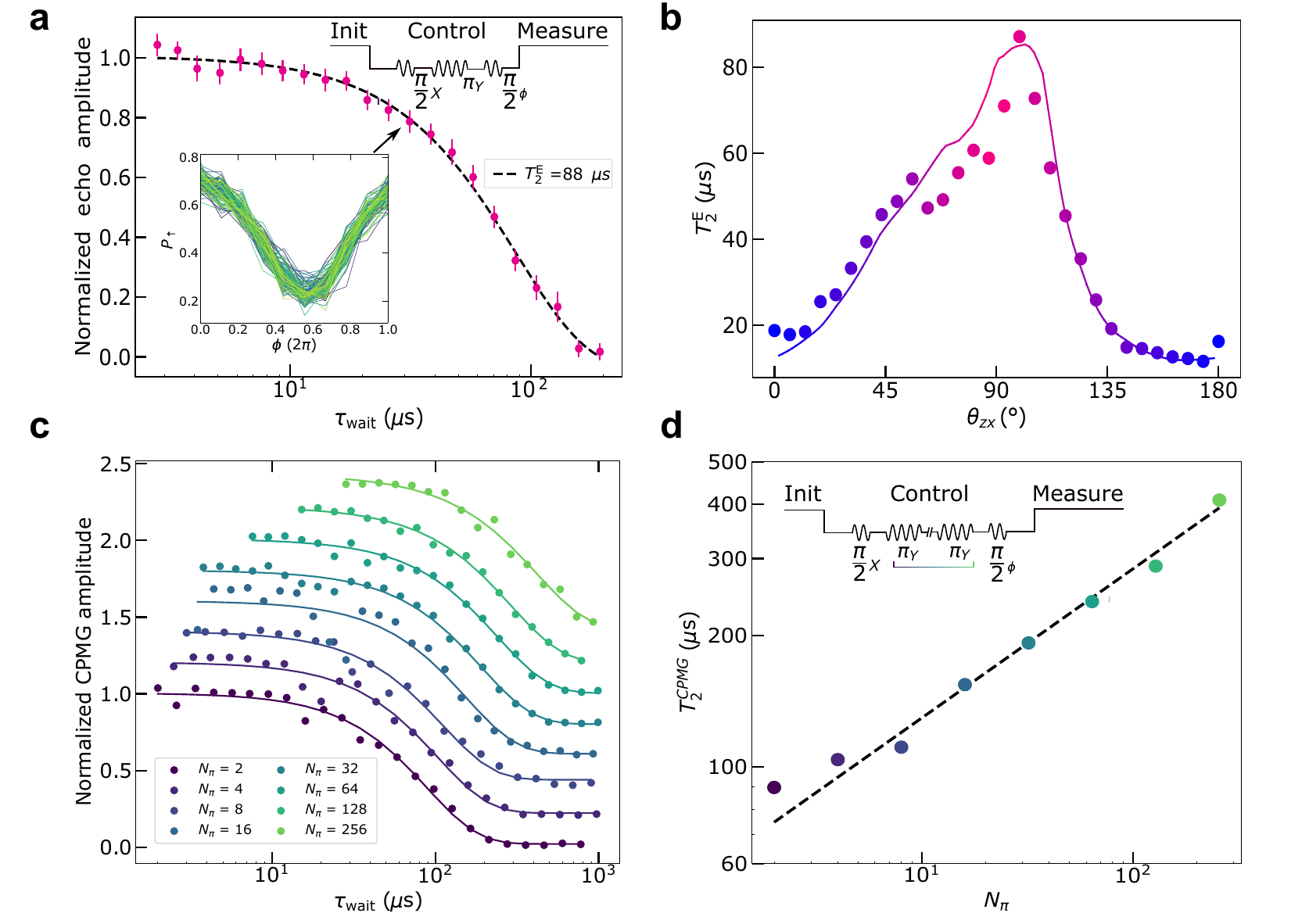}
\caption{\label{fig3} \textbf{Anisotropy of the hole spin coherence and sweet-spot operation.} \textbf{(a)} Normalized Hahn-echo amplitude {\it vs} free evolution time $\tau_{\rm wait}$ at $f_L=17$\,GHz. The top-right inset sketches the pulse sequence. The bottom-left inset displays $P_{\uparrow}(\tau_{\rm wait}=31.4\,\mu{\rm s})$ {\it vs} the phase $\phi$ of the last $\pi/2$ pulse for 100 repetitions. For each $\tau_{\rm wait}$, we extract the average amplitude of the $P_{\uparrow}(\phi)$ oscillations and normalize it to the average amplitude in the zero-delay limit. The resulting normalized echo amplitudes are reported on the main plot. The dashed curve is a fit to $\exp(-(\tau_{\rm wait}/{T_2^E})^{\beta})$ with $\beta = 1.5\pm 0.1$. \textbf{(b)} Measured $T_2^{\rm E}$ vs magnetic field angle $\theta_{zx}$ (symbols). The solid line is a fit to Eq.~(\ref{sigmas}), using the experimental $\LSESone$ and $\LSEStwo$ from Figs.~\ref{fig2}a and \ref{fig2}d. \textbf{(c)} Normalized CPMG amplitude as a function of free evolution time $\tau_{\rm wait}$ for different numbers $N_\pi$ of $\pi$ pulses (curves are offset for clarity). The solid lines are fits to the same exponential decay function as in (a) with $\beta=1.5$. \textbf{(d)} Extracted $T_2^{\rm CPMG}$ as a function of $N_\pi$. The dashed line is a linear fit with slope $\gamma = 0.34$. The inset sketches the CPMG pulse sequence: $N_\pi$ equally spaced $\pi_y$ pulses between two $\pi_x/2$ pulses. As for Hahn-echo, we detune the phase of the last pulse.}
\end{figure*}

\section*{Coherence times}

We now turn to the angular dependence of the hole spin coherence time and investigate its correlation with the longitudinal spin-electric susceptibility \cite{Tanttu_PRXQ_2019}. 
To get rid of low frequency noise sources, we measure the coherence time using a conventional Hahn-echo protocol \cite{burkard2021}. The control sequence, applied to  \Gone{} (see upper inset of Fig.~\ref{fig3}a), consists of $\pi_x/2$, $\pi_y$ and $\pi_\phi/2$ pulses separated by a time delay $\tau_{\rm wait}/2$. For each $\tau_{\rm wait}$, we extract the averaged amplitude of the $P_{\uparrow}$ oscillation obtained by varying the phase $\phi$ of the last $\pi/2$ pulse, and normalize it to the $P_{\uparrow}$ oscillation amplitude in the zero-delay limit. 

A representative Hahn-echo plot is shown in Fig.~\ref{fig3}a. We fit the echo amplitude to an exponential decay $\exp(-(\tau_{\rm wait}/T_2^{\rm E})^{\beta})$, where the  exponent $\beta$ is left as a free parameter. The best fit is obtained for $\beta = 1.5\pm 0.1$, which implies a high frequency noise with a characteristic spectrum $S(f)=S_{\rm hf}(f_0/f)^\alpha$, where $f_0=1$\,Hz is a reference frequency and $\alpha=\beta-1\approx 0.5$ (we note that the same $\alpha$ value was reported for hole spin qubits in germanium \cite{Veldhorst_2021}).

To explore the angular dependence of $T_2^{\rm E}$ in the $xz$ plane, we measure the decay of the Hahn-echo amplitude for different values of $\theta_{zx}$. The results, shown in Fig.~\ref{fig3}b, reveal a strong anisotropy, with $T_2^{\rm E}$ ranging from $15$\,$\mu$s to $88$\,$\mu$s. Strikingly, the spin coherence time peaks at $\theta_{zx}=99\degree$, an angle between the minimum of $|\LSESone|$ and a zero of $\LSEStwo$, highlighting a correlation with the correspondingly suppressed electrical noise. The extended coherence time is much longer than previously reported for hole spin qubits in both silicon and germanium \cite{stano_2021_review}. In addition, we notice that spin control remains efficient at all angles including $\theta_{zx}=99\degree$, where we could readily achieve Rabi frequencies $F_{\rm Rabi}$ as large as 5 MHz limited by the attenuation on the microwave line. The echo quality factor $Q^{\rm E}=F_{\rm Rabi}\times T_2^{\rm E}$ also peaks at $\theta_{zx}=99\degree$, reaching $Q^{\rm E}\approx 440$ with further room for improvement (see Supp. Infos S4 and S5).

The observed angular dependence of $T_2^{\rm E}$ can be understood by assuming that the electrical noise is the sum of uncorrelated voltage fluctuations on the different gates G$i$ with respective spectral densities $S_{{\rm G}i}(f)=\Shf (f_0/f)^{0.5}$. Given the Hahn-Echo noise filter function, the decoherence rate can then be expressed as (see Supp. Info S6):
\begin{equation}
\label{sigmas}
\frac{1}{T_2^{\rm E}}\approx 7.8f_0^{1/3}\left(\sum_{i}\left(\frac{\partial f_L}{\partial V_{{\rm G}i}}\right)^{2}\Shf\right)^{2/3}\,. 
\end{equation}
Using the longitudinal spin-electric susceptibilities from Figs.~\ref{fig2}a and \ref{fig2}d and leaving the weights $\Shf$ as adjustable parameters, we achieve a remarkable agreement with the experimental $T_2^{\rm E}$ (see colored solid line in Fig.~\ref{fig3}b). 
This strongly supports the hypothesis that the Hahn-echo coherence time is limited by electrical noise. As already argued before, $\LSESone{}$ and $\LSEStwo{}$ indeed quantify the susceptibility of the hole spin to electric field fluctuations parallel and perpendicular to the channel, respectively. 

The best fit in Fig.~\ref{fig3}b is obtained with $\Shfi{1}=( 1.7 \,\mu{\rm V/\sqrt{Hz}})^2$ and $\Shfi{2}=( 66 \,n{\rm V/\sqrt{Hz}})^2$.
We speculate that the large $\Shfi{1}/\Shfi{2}$ ratio results from an artificial enhancement of $\Shfi{1}$ accounting for hidden sources of electric field fluctuations along the silicon nanowire. Certainly, Eq.~(\ref{sigmas}) misses the contribution from the electrical noise on \Gthree{}, whose LSES could not be measured. For a symmetry reason, we expect $\LSESthree$ to be comparable to $\LSESone$. A possible additional source of longitudinal electric field fluctuations is the randomly oscillating charges and dipoles in the silicon nitride spacers between the gates. Because these noise sources are closer to \Qtwo{} than is gate \Gone{}, and because they are much less screened by the hole gas beneath, they presumably make a large contribution to the apparent $\Shfi{1}$ when lumped into $\propto \LSESone$ terms.

To further investigate the hole spin coherence, we implement Carr–Purcell–Meiboom–Gill (CPMG) sequences at the most favorable field orientation $\theta_{zx}=99\degree$. These consist in increasing the number of $\pi$ pulses cancelling faster and faster dephasing mechanisms. Figure~\ref{fig3}c displays the CPMG echo amplitudes as a function of the total waiting time $\tau_{\rm wait}$ for series of $N_\pi=2^n$ $\pi$ pulses, where $n$ is an integer ranging from $1$ to $8$. The CPMG decay times $T_2^{\rm CPMG}$ extracted from Fig.~\ref{fig3}c (see caption) are plotted against $N_\pi$ in Fig.~\ref{fig3}d. As expected, the data points follow a power law $T_2^{\rm CPMG}\propto N_{\pi}^\gamma$, where $\gamma=\frac{\alpha}{\alpha+1}$ for a $\propto 1/f^{\alpha}$ noise spectrum \cite{Yoneda-2018}. The best fit value $\gamma=0.34$ yields again $\alpha\approx0.5$. For the largest sequence of 256 $\pi$ pulses, we find $T_2^{\rm CPMG}=0.4$\ ms, which is the longest coherence ever reported for hole spins \cite{stano_2021_review}. 

\begin{figure*}[h!]
\includegraphics[width = 1 \textwidth]{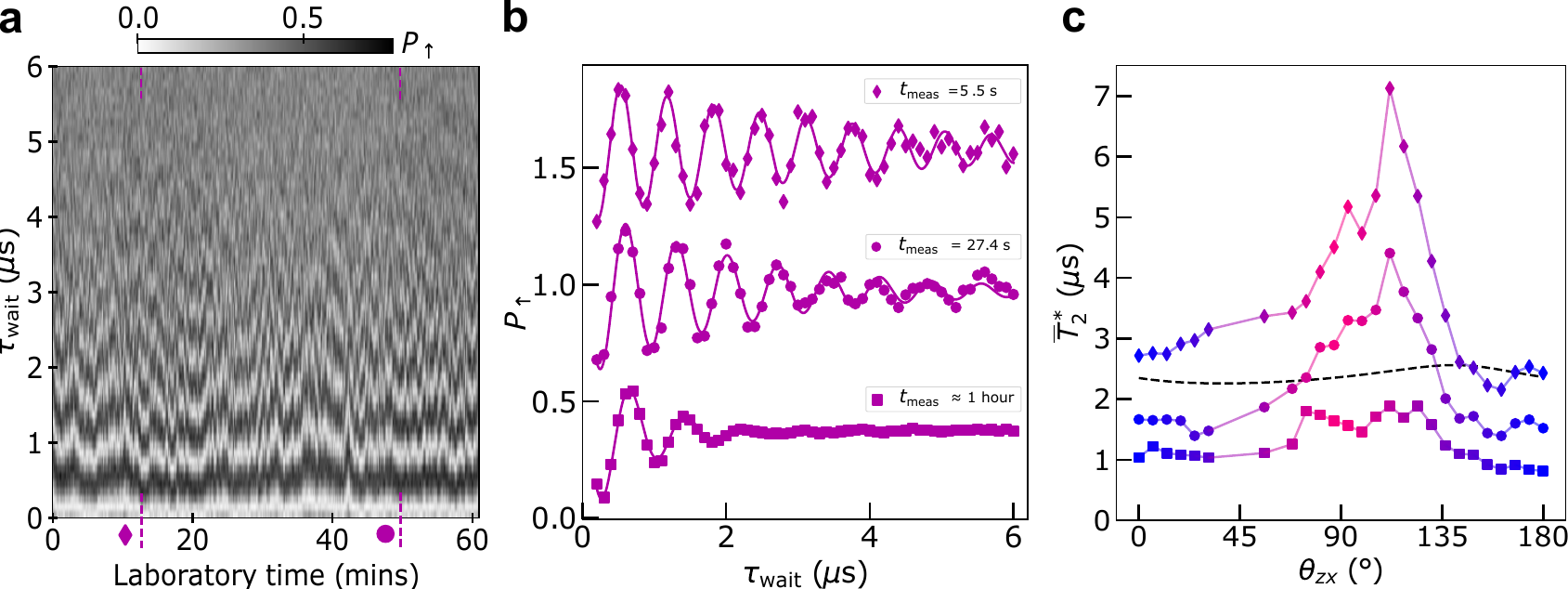}
\caption{\label{fig4} \textbf{Free induction decay (FID).} \textbf{(a)} Collection of 600 Ramsey oscillations as a function of $\tau_{\rm wait}$, the free evolution time between two $\pi_x/2$ pulses, at $\theta_{zx}=118\degree$. The applied microwave frequency is detuned by $\sim$ 700 kHz from the Larmor frequency. Each Ramsey oscillation is measured in $\approx 5.5$\,s. The locations of the representatives traces shown in (b) are indicated by a diamond and a dot. \textbf{(b)} Selected averages of Ramsey oscillations taken over different measurement times: $t_{\rm meas}=5.5$\,s corresponding to a single trace (diamonds); $t_{\rm meas}=27.5$\,s, corresponding to 5 consecutive traces (circles); $t_{\rm meas}\approx 1$\,hour, corresponding to the full set of 600 traces (squares). The solid lines are fits to Gaussian decaying oscillations. Note that the decay time $T_2^*$ depends on the chosen subset of consecutive traces (except for $t_{\rm meas}\approx 1$\,hour), which is a signature of non-ergodicity\cite{Delbecq_PRL_2016} at small $t_{\rm meas}$ (see Supp. Info S7). Hence we observe a distribution of $T_2^*$ values with mean $\overline{T}_2^*$. \textbf{(c)} Mean $\overline{T}_2^*$ for different $t_{\rm meas}$ (same symbols as in (b)) as a function of the magnetic field angle $\theta_{zx}$. The solid lines are guides to the eye. The dashed black line is the calculated dephasing time due to hyperfine interactions (see Supp. Info S8). 
}
\end{figure*}

Finally, to gain insight into the low frequency noise acting on the hole spin, we perform systematic measurements of the inhomogeneous dephasing time $T_2^*$. To this aim, we apply Ramsey control sequences consisting of two $\pi/2$ pulses separated by a variable delay $\tau_{\rm wait}$. Contrary to Hahn-echo, the dephasing induced by low frequency noise sources is not cancelled due to the absence of the refocusing $\pi$ pulse. Figure~\ref{fig4}a displays $P_\uparrow$ for a series of identical Ramsey sequences recorded on an overall time frame of one hour, with each sequence lasting approximately 5.5\,s. The next step is to average $P_\uparrow(\tau_{\rm wait})$ on a subset of consecutive sequences measured within a total time $t_{\rm meas}$. This way, an averaged Ramsey oscillation is obtained for each $t_{\rm meas}$, whose amplitude is fitted to a Gaussian-decay function yielding $T_2^*(t_{\rm{meas}})$. Representative Ramsey data sets and corresponding fits are shown in Fig.~\ref{fig4}b for three values of $t_{\rm meas}$. The inhomogeneous dephasing time decreases with increasing $t_{\rm meas}$ due to the contribution of noise components with lower and lower frequency. To unveil the angular dependence of $T_2^*$, we repeat the same measurement for different magnetic field orientations. The results are plotted in Fig.~\ref{fig4}c for the same three values of $t_{\rm meas}$. The overall anisotropy of the Hahn-echo decay time of Fig.~\ref{fig3}b can still be identified, although it reduces at large $t_{\rm meas}$ starting from $t_{\rm meas}>50$\,s.

However, if the $1/f^{0.5}$ charge noise prevailed over the whole mHz to MHz range, $T_2^*$ would be $\approx 50$\,$\mu$s when $T_2^{\rm E}\approx88\,\mu$s (see Supp. Info S6), well above the $7\,\mu$s seen in Fig.~\ref{fig4}c. The power spectrum $S(f)$ at low frequency can be extracted from the data of Fig.~\ref{fig4}a (see Supp. Info S9). This reveals a $1/f^\alpha$ noise with $\alpha$ closer to 1, and a power (at 1 Hz) four orders of magnitude larger than the one expected by extrapolating the high-frequency $1/f^{0.5}$ noise inferred from CPMG. The change of color and amplitude of $S(f)$ when going from the mHz to the MHz points to the presence of different mechanisms dominating the dephasing at low and high frequencies. 

We note that the $T_2^*\approx 1-2$\,$\mu$s measured at long $t_{\rm meas}$ is below but fairly close to the expected hole spin dephasing time due to hyperfine interactions with the naturally present $^{29}$Si nuclear spins \cite{Voisin_NL_2014} (see the dashed line in Fig.~\ref{fig4}c and Supp. Info S8 for details). This suggests that low-frequency dephasing may be partially due to such hyperfine interactions. 

In conclusion, we have reported on the first spin qubit with electrical control and single-shot readout based on a single hole in a silicon nanowire device issued from an industrial-grade fabrication line. The hole wave function and corresponding $g$-factors could be modeled with an unprecedented level of accuracy in these types of devices, denoting a relatively low level of structural and charge disorder.
The hole-spin coherence was found to be limited by a $1/f^{0.5}$ charge noise at high frequencies ($10^4-10^6$ Hz), with a strong dependence on the magnetic-field orientation that could be faithfully accounted for by the spin-electric susceptibilities. A largely enhanced spin coherence was measured at the sweet-spot angle, far beyond the current state-of-the-art for hole-spin qubits and close to the best figures reported for $^{28}$Si electron-spin qubits electrically driven via a micro-magnet. Our study of the inhomogeneous dephasing time revealed a much stronger noise at low frequencies ($10^{-4}-10^{-2}$ Hz) that could be partially ascribed to the expected hyperfine interaction. In this scenario, the possible introduction of isotopically purified silicon devices would lead to significant improvement of hole-spin coherence in the low-frequency range.  
Finally, we would like to emphasize that such sweet spots shall be ubiquitous in hole spin qubit devices\cite{Bosco_PRXQ_2021}, and that a careful design and choice of operation point can make them pretty robust to disorder (see example in Supp. Info S2). The engineering of sweet spots shall therefore open new opportunities for an efficient realization of multi-qubit or coupled spin-photon systems\cite{Michal2022}.


\small{
\subsection*{Methods}

\textbf{Device.}

The device is a four-gate silicon-on-insulator nanowire transistor fabricated in an industry-standard 300-mm CMOS platform  \cite{Maurand_Ncomms_2016}. The undoped [110]-oriented silicon nanowire channel is $17$\,nm thick and $100$\,nm wide. It is connected to wider boron-doped source and drain pads used as reservoirs of holes. The four wrapping gates  (\Gone{}, \Gtwo{}, \Gthree{} and \Gfour{}) are $40$\,nm long and they are spaced by $40$\,nm. The gaps between adjacent gates and between the outer gates and the doped contacts are filled by silicon nitride spacers. The gate stack consists of a $6$ nm thick SiO$_2$ dielectric layer followed by a metallic bilayer with 6 nm of TiN and 50 nm of heavily doped poly-silicon. The yield of the 4-gate devices across the full 300 mm wafer reaches 90\% and their room temperature characteristics exhibit excellent uniformity (see Supp. Info S10 for details).

\textbf{Dispersive readout.}

Similar to charge detection methods recently applied to SOI nanowire devices \cite{Chanrion2020,Ansaloni2020}, we accumulate a large hole island under the gates \Gthree{} and \Gfour{}, as sketched in Fig.~\ref{fig1}a. The island acts both as a charge reservoir and electrometer for the quantum dot \Qtwo{} located under \Gtwo{}. However, unlike the above-mentioned earlier implementations, the electrometer is sensed by rf dispersive reflectometry on a tank LC resonator connected to the drain rather than to a gate electrode. To this aim, a commercial surface-mount inductor ($L=240$\,nH) is wire bonded to the drain pad. This configuration involves a parasitic capacitance to ground $C_p=0.54$\,pF, leading to resonance frequency $f=449.81$\,MHz. The high value of the loaded quality factor $Q \approx 10^3$ enables fast, high-fidelity charge sensing. We estimate a charge readout fidelity of 99.6\% in 5 $\mu$s, which is close to the state-of-the-art for Si MOS devices \cite{Schaal2020}. The resonator characteristic frequency experiences a shift at each Coulomb resonance of the hole island, \textit{i.e.} when the electrochemical potential of the island lines up with the drain Fermi energy. This leads to a dispersive shift in the phase $\phid$ of the reflected radio-frequency signal, which is measured through homodyne detection (See Supp. Info S1 for details on the spin readout and S11 for measurement setup).

\textbf{Pulse sequences.}

For Ramsey, Hahn echo, phase gate and CPMG pulse sequences, we set a $\pi/2$ rotation time of 50~ns. Given the angular dependence of $F_{\rm Rabi}$, we calibrate the microwave power required for this operation time for each magnetic field orientation. We also calibrate the amplitude of the $\pi$ pulses to achieve a $\pi$ rotation in 150~ns.
In extracting the noise exponent $\gamma$ from CPMG measurements, we do not include the time spent in the $\pi$ pulses (this time amounts to about 10 \% of the duration of each pulse sequence).

\textbf{Modeling.}
The hole wave functions and $g$-factors are calculated with a 6-band $\mathbf{k}\cdot\mathbf{p}$ model\cite{Venitucci_PRB_2018}. The screening by the hole gases under gates \Gone{}, \Gthree{} and \Gfour{} is accounted for in the Thomas-Fermi approximation. As discussed extensively in Supp. Info S2, the best agreement with the experimental data is achieved by introducing a moderate amount of charge disorder. The theoretical data displayed in Figs. \ref{fig1} and \ref{fig2} correspond to a particular realization of this charge disorder (point-like positive charges with density $\sigma=5\times 10^{10}$ cm$^{-2}$ at the Si/SiO$_2$ interface and $\rho=5\times 10^{17}$ cm$^{-3}$ in bulk Si$_3$N$_4$). The resulting variability is discussed in Supp. Info S2. The rotation of the principal axes of the $g$-tensor are most likely due to small inhomogeneous strains ($<0.1\%$); however, in the absence of quantitative strain measurements, we have simply shifted $\theta_{zx}$ by $\approx -25\degree$ and $\theta_{zy}$ by $\approx 10\degree$ in the calculations of Figs. \ref{fig1} and \ref{fig2}.

\textbf{Data availability.} 

All of the data used to produce the figures in this paper and to support our analysis and conclusions are available upon reasonable request to the corresponding author.

\subsection*{Acknowledgments}

This research has been supported by the European Union’s Horizon 2020 research and innovation programme under grant agreements No. 951852 (QLSI project), No. 810504 (ERC project QuCube) and  No. 759388 (ERC project LONGSPIN), and by the French National Research Agency (ANR) through the projects MAQSi and CMOSQSPIN. 

\subsection*{Authors contributions}
N.P. and B.Br. carried out the experiment with help from V.S., S.Z., and A.A. and under the supervision of X.J., R.M. and S.D. V.M, J.C.A.U. and Y.M.N. carried out the theoretical modeling. B.Be,  H. N., L.H. and M.V. designed and supervised the fabrication of the device. M.U. and T.M. provided useful comments. N.P., B.Br., Y.M.N., R.M. and S.D co-wrote the papers with input from the other authors.    
\subsection*{Competing financial interests}

The authors declare no competing financial interests.
}

\clearpage

\setcounter{section}{0}
\setcounter{equation}{0}
\setcounter{figure}{0}
\setcounter{table}{0}

\renewcommand\thefigure{S.\arabic{figure}} 
\renewcommand\thesection{S\arabic{section}} 
\renewcommand{\theHfigure}{S.\arabic{figure}}

\begin{center}
\textbf{\large Supplementary information for ``A single hole spin with enhanced coherence in natural silicon''}
\end{center}


\section{\label{suppinf:readout} Energy selective single shot readout of spin state of the first hole in \Qtwo}

\begin{figure*}[h]
\includegraphics[width = 0.95\textwidth]{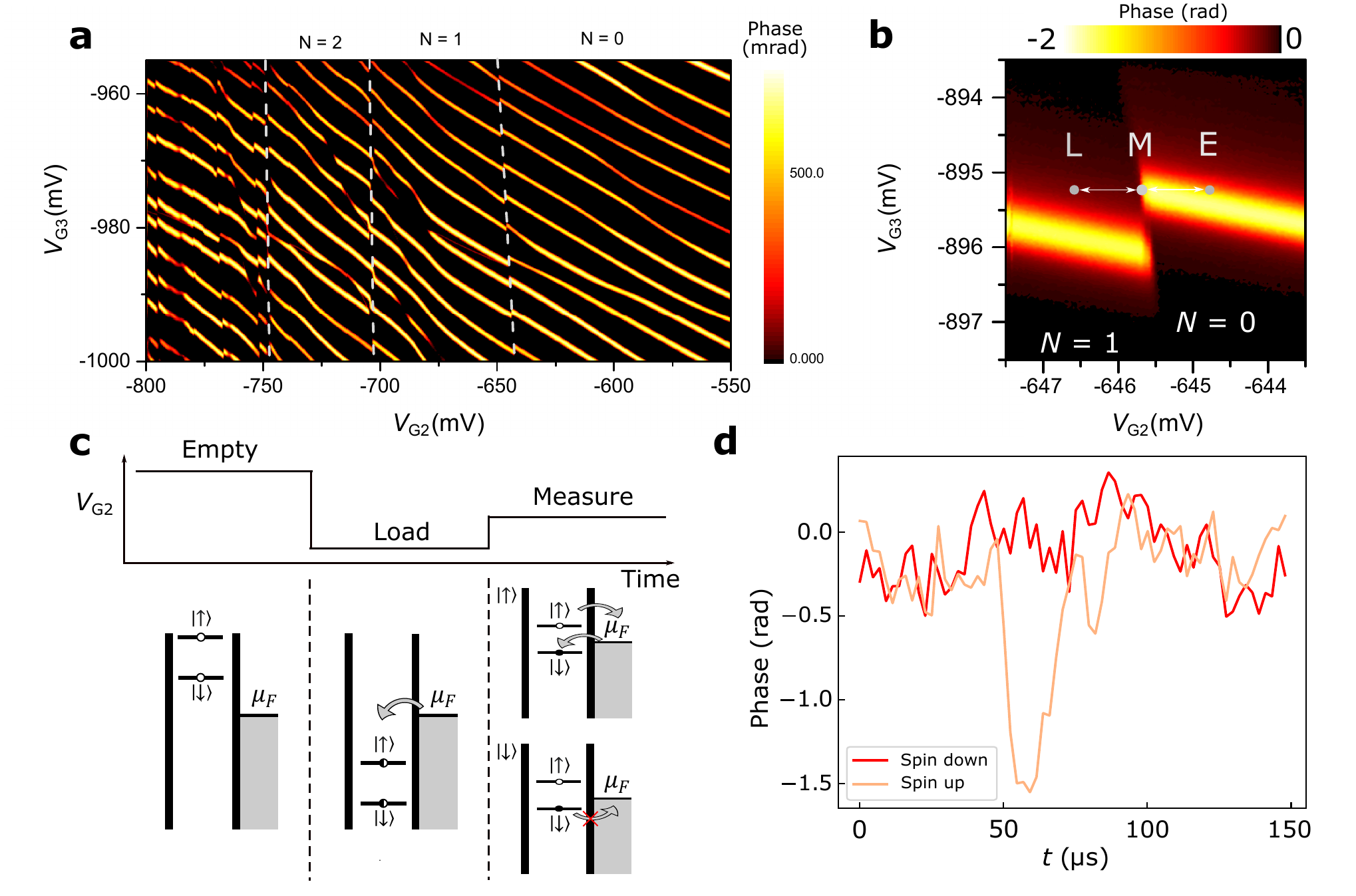}
\caption{\textbf{Single shot spin readout.} \textbf{(a)} Stability diagram of the device as a function of $\Vgtwo$ and $\Vgthree$. The dashed grey lines are guides to the eye highlighting charge transitions in \Qtwo{}. The first hole tunnels into \Qtwo{} at $\Vgtwo\approx -650$\,mV. \textbf{(b)} Zoom on the stability diagram close to the working point used in the main text. The points labelled L (Load), M (Measure) and E (Empty) are the three stages of the pulse sequence applied to $V_{\rm G2}$ for spin readout. \textbf{(c)} (Top) Schematic of the three stages pulse sequence applied to $\Vgtwo$. (Bottom) Schematic energy diagrams at the different stages of the pulse sequence. $\mu_F$ is the chemical potential of the charge sensor playing the role of reservoir. A random spin is charged during the load stage. At the measure stage, if the loaded spin is up, the hole is able to tunnel out and is replaced by a spin down. On the opposite, if the loaded spin is down, tunneling in or out is impossible. Finally, the dot is discharged during the empty stage. \textbf{(d)} Phase versus time during the measurement stage. The orange curve exhibits a ``blip'' around $t=50\,\mu$s, which indicates that the dot experienced a discharge/charge cycle characteristic of a spin up loading (see c). On the contrary, the red curve shows no phase change, which can be interpreted as a spin down loading. The phase signal is integrated over $6\,\mu$s.}
\label{fig:readout}
\end{figure*}

Figure~\ref{fig:readout}a displays the stability diagram of the device as a function of $\Vgtwo$ and $\Vgthree$ when a large quantum dot (acting as a charge sensor) is accumulated under gates G3 and G4. The dashed grey lines outline the charging events in the quantum dot \Qtwo{} under G2, detected as discontinuities in the Coulomb peak stripes of the sensor dot. The lever-arm parameter of gate G2 is $\alpha\approx 0.37$ eV/V, as inferred from temperature-dependence measurements. Comparatively, the lever-arm parameter of gate G1 with respect to the first hole under \Gtwo{} is $\alpha_{G1}\approx 0.03$ eV/V.
The charging energy, measured as the splitting between the first two charges is $U=22$ meV. Figure~\ref{fig:readout}b shows a zoom on the stability diagram around the working point used for single shot spin readout in the main text. The three points labelled Empty (E), Load (L) and Measure (M) are the successive stages of the readout sequence sketched in Fig.~\ref{fig:readout}c. The quantum dot is initially emptied (E) before loading (L) a hole with a random spin. Both spin states are separated by the Zeeman energy $E_Z=g\mu_B B$ where $g$ is the $g$-factor, $\mu_B$ the Bohr magneton and $B$ the amplitude of the magnetic field. This opens a narrow window for energy selective readout using spin to charge conversion \cite{Elzerman_Nature_2004}. Namely, we align at stage (M) the center of the Zeeman splitted energy levels in \Qtwo{} with the chemical potential of the sensor. In this configuration, only the excited spin up hole can tunnel out of \Qtwo{} while only spin down holes from the sensor can tunnel in. These tunneling events are detected by thresholding the phase of the reflectometry signal of the sensor to achieve single shot readout of the spin state. Typical time traces of the reflected signal phase at stage (M), representative of a spin up (spin down) in \Qtwo{}, are shown in Fig.~ \ref{fig:readout}d.

We used this three stage pulse sequence to optimize the readout. For that purpose, the tunnel rates between \Qtwo{} and the charge sensor were adjusted by fine tuning $\Vgthree{}$ and $\Vgfour{}$. For the spin manipulation experiment discussed in the main text, we used a simplified two stages sequence for readout by removing the empty stage. The measure stage duration was set to $200\,\mu$s for all experiments, while the load stage duration (seen as a manipulation stage duration) was ranging from $50\,\mu$s to $1$ ms. In order to obtain the spin-up probability $P_\uparrow$ after a given spin manipulation sequence, we repeated the single-shot readout a large number of times, typically 100 to 1000 times.

\section{\label{suppinf:Modeling}Modeling of the $g$-factors}

In this section, we give an overview of the methodology used to model the device, then discuss the outcome of the simulations and the comparison with experimental data. We finally provide arguments on the robustness of the sweet spots as an outlook.

\subsection{Methodology}

\begin{figure*}[h!]
\includegraphics[width = 0.5\textwidth]{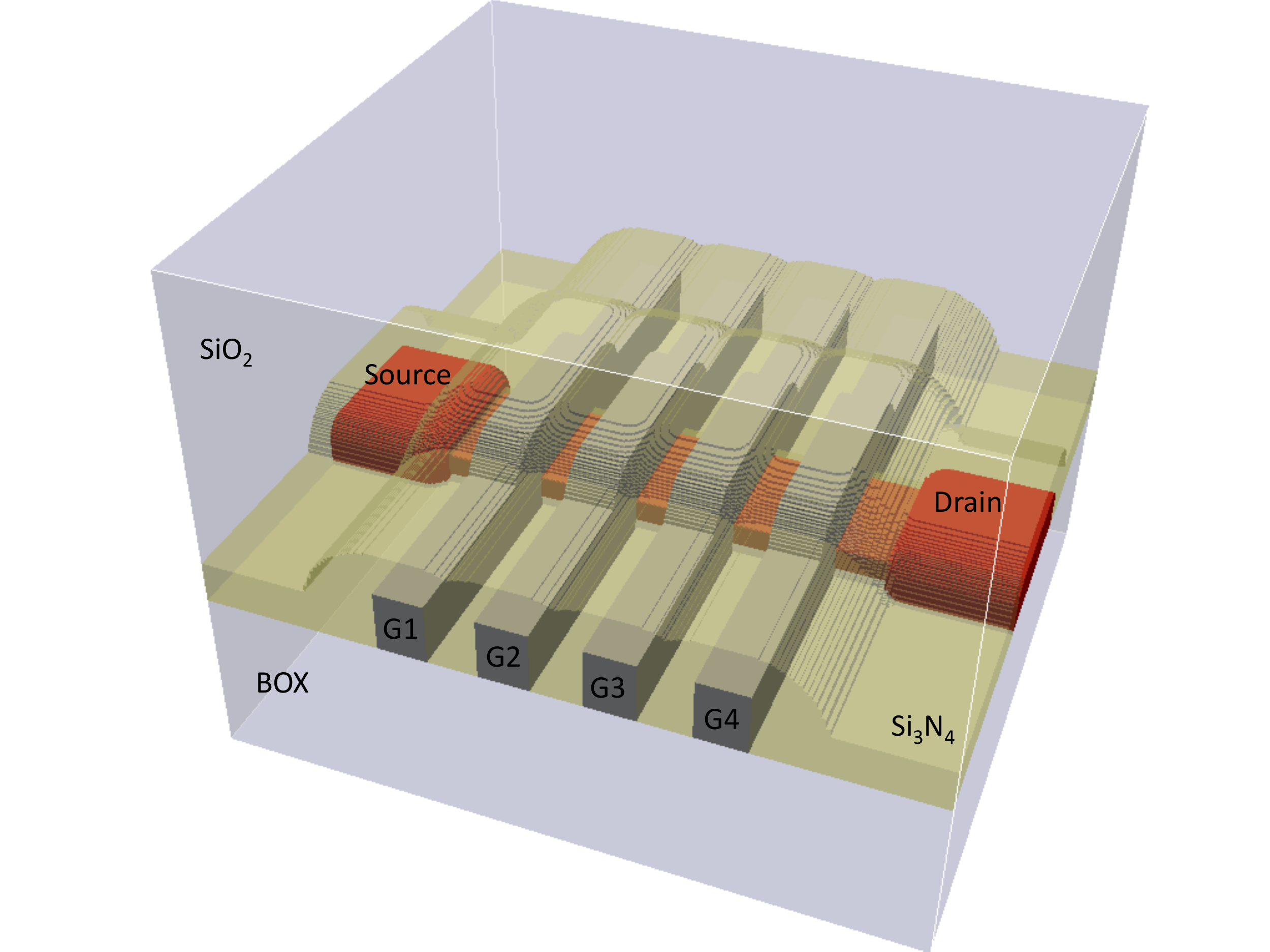}
\caption{\textbf{Modeled structure.} The 17 nm thick and 100 nm wide silicon channel is connected to highly doped source and drain reservoirs and controlled by four gates \Gone{}...\Gfour{}. The substrate below the BOX and the top gate above the structure (at the Metal 1 level) are grounded.}
\label{fig:DeviceModel}
\end{figure*}

The device (Fig. \ref{fig:DeviceModel}) is modeled as a $[110]$-oriented rectangular nanowire channel with width $W=100$ nm and height $H=17$ nm lying on a 145 nm thick buried oxide (BOX). Four 40 nm long and 50 nm tall front gates, separated by 40 nm long Si$_3$N$_4$ spacers, are laid across the channel. They are insulated by a 6 nm thick SiO$_2$ layer. Highly doped source and drain reservoirs ($N_{\rm A}=10^{20}$ cm$^{-3}$) are overgrown at both ends of the channel. The whole device is embedded in a 35 nm thick Si$_3$N$_4$ contact etch stop layer (CESL), and coated with a $\simeq 250$ nm thick oxide. The silicon substrate beneath can be used as a back gate, and a wide metal line above (at the Metal 1 level) as an extra top gate. These top and back gates, as well as the source and drain are grounded in the simulations.

The potential landscape $V(\mathbf{r})$ in the device is first computed with a finite volumes Poisson solver \cite{Venitucci_PRB_2018}. Screening by the holes accumulated in the source, drain and below the gates \Gone{}, \Gthree{}, and \Gfour{} is accounted for in the Thomas-Fermi approximation. Namely, these accumulations are modeled as locally homogeneous 3D hole gases, with density:
\begin{equation}
p(\mathbf{r})=N_v F_{1/2}\left[\beta\left(E_v-e V(\mathbf{r})-\mu\right)\right]\,,
\end{equation}
where $F_{1/2}$ is a Fermi-Dirac integral, $N_v=(3.5\times 10^{15}\,{\rm cm}^{-3}.{\rm K}^{-3/2})T^{3/2}$ is the effective density of states in the valence band, $E_v-\mu$ is difference between the valence band edge energy and the chemical potential (chosen to match the threshold voltage of the device), and $\beta=1/k_BT$ with $T$ the temperature. This equation is solved self-consistently together with Poisson's equation:
\begin{equation}
\varepsilon_0\mathbf{\nabla}\varepsilon_r(\mathbf{r})\cdot\mathbf{\nabla} V(\mathbf{r})=- e\left[p(\mathbf{r})+\rho_{\rm test}(\mathbf{r})+\rho_{\rm trap}(\mathbf{r})-N_{\rm A}(\mathbf{r})\right]\,,
\end{equation}
where $\varepsilon_0\varepsilon_r(\mathbf{r})$ is the material-dependent dielectric constant, $\rho_{\rm test}(\mathbf{r})$ is a test charge distribution that mimics a single hole within the dot \Qtwo{} under \Gtwo, and $\rho_{\rm trap}(\mathbf{r})$ is a distribution of charge traps used to assess the effects of disorder. The test charge $\rho_{\rm test}(\mathbf{r})$ prevents the Thomas-Fermi density from flooding the dot, as this approximation is notoriously inaccurate in the few holes regime. The bias voltages are used to set the boundary conditions on the gates. 

The test charge distribution $\rho_{\rm test}(\mathbf{r})$ is practically modeled as a homogeneous ellipsoid with total charge $+1$, centered on the average position $\mathbf{R}=(\langle x\rangle, \langle y\rangle, \langle z\rangle)$ of the hole (computed {\it a posteriori} from the quantum-mechanical wave functions), with radii $a_x=\sqrt{3(\langle x^2\rangle-\langle x\rangle^2)}$, $a_y=\sqrt{3(\langle y^2\rangle-\langle y\rangle^2)}$, and $a_z=\sqrt{3(\langle z^2\rangle-\langle z\rangle^2)}$. As the potential $V_{\rm QD}(\mathbf{r})$ relevant for the Hamiltonian of the dot is that of the empty \Qtwo, the self-consistent $V(\mathbf{r})$ is corrected from the contribution of $\rho_{\rm test}(\mathbf{r})$:
\begin{equation}
V_{\rm QD}(\mathbf{r})=V(\mathbf{r})-V_{\rm test}(\mathbf{r})\,,
\end{equation}
where $V_{\rm test}(\mathbf{r})$ is the potential created by $\rho_{\rm test}(\mathbf{r})$:
\begin{equation}
\varepsilon_0\mathbf{\nabla}\varepsilon_r(\mathbf{r})\cdot\mathbf{\nabla} V_{\rm test}(\mathbf{r})=-e\rho_{\rm test}(\mathbf{r})\,.
\end{equation}
As long as the dot and hole gases around remain sufficiently separated, the resulting $V_{\rm QD}(\mathbf{r})$ is only weakly dependent on the choice of $\rho_{\rm test}(\mathbf{r})$.

The wave functions in the potential $V_{\rm QD}(\mathbf{r})$ are then calculated on the same mesh with a finite differences 6 bands $\mathbf{k}\cdot\mathbf{p}$ model \cite{Venitucci_PRB_2018}. We use Luttinger parameters $\gamma_1=4.285$, $\gamma_2=0.339$, $\gamma_3=1.446$, split-off energy $\Delta=44$ meV and Zeeman parameter $\kappa=-0.42$. The $g$-matrix of the ground-state is finally computed along the lines of Ref. \citenum{Venitucci_PRB_2018}. The present formalism captures all the effects of spin-orbit coupling, including Rashba-type interactions when the dot moves along the channel \cite{Crippa_PRL_2018,Michal_PRB_2021}.

\subsection{Discussion}

In the following, we first discuss the nature of the hole states, and show that they tend to be confined in the top left or right corners of the channel by the lateral component of the electric field of the non-planar gate. We then argue why charge disorder needs to be introduced to reach lateral electric fields compatible with the experimental data. We discuss the resulting variability of the $g$-factors. Finally, we identify strain as the most likely mechanism for the rotation of the principal axes of the $g$-tensor evidenced on Fig. 1 of the main text.

\subsubsection{Nature of the hole states.}

If the device were ``planar'', the hole would be confined at the top $(001)$ facet of the channel by the quasi-vertical electric field of gate \Gtwo{}. It would, therefore, show the fingerprints of an almost pure $(001)$ heavy-hole, with a large $g_x\simeq-6\kappa+2\gamma_h\simeq 4.84$, and much smaller $g_y$ and $g_z$ characteristic of the weak heavy-hole/light-hole mixing induced by the lateral confinement ($\gamma_h=1.16$ being a correction that describes the heavy-hole/light-hole mixing by the magnetic vector potential) \cite{Michal_PRB_2021}.

\begin{figure*}[t]
\includegraphics[width = 0.475\textwidth]{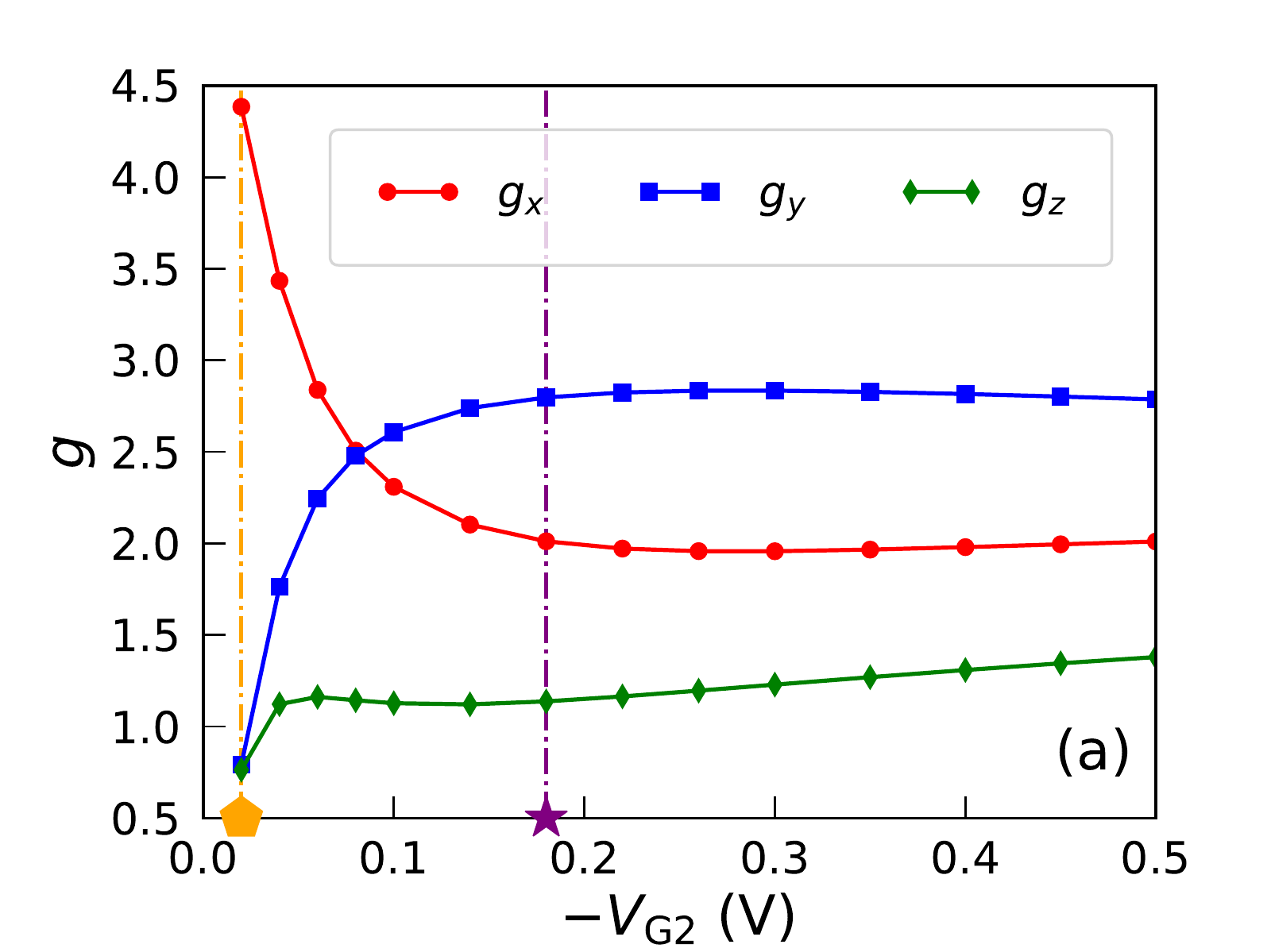}
\includegraphics[width = 0.475\textwidth]{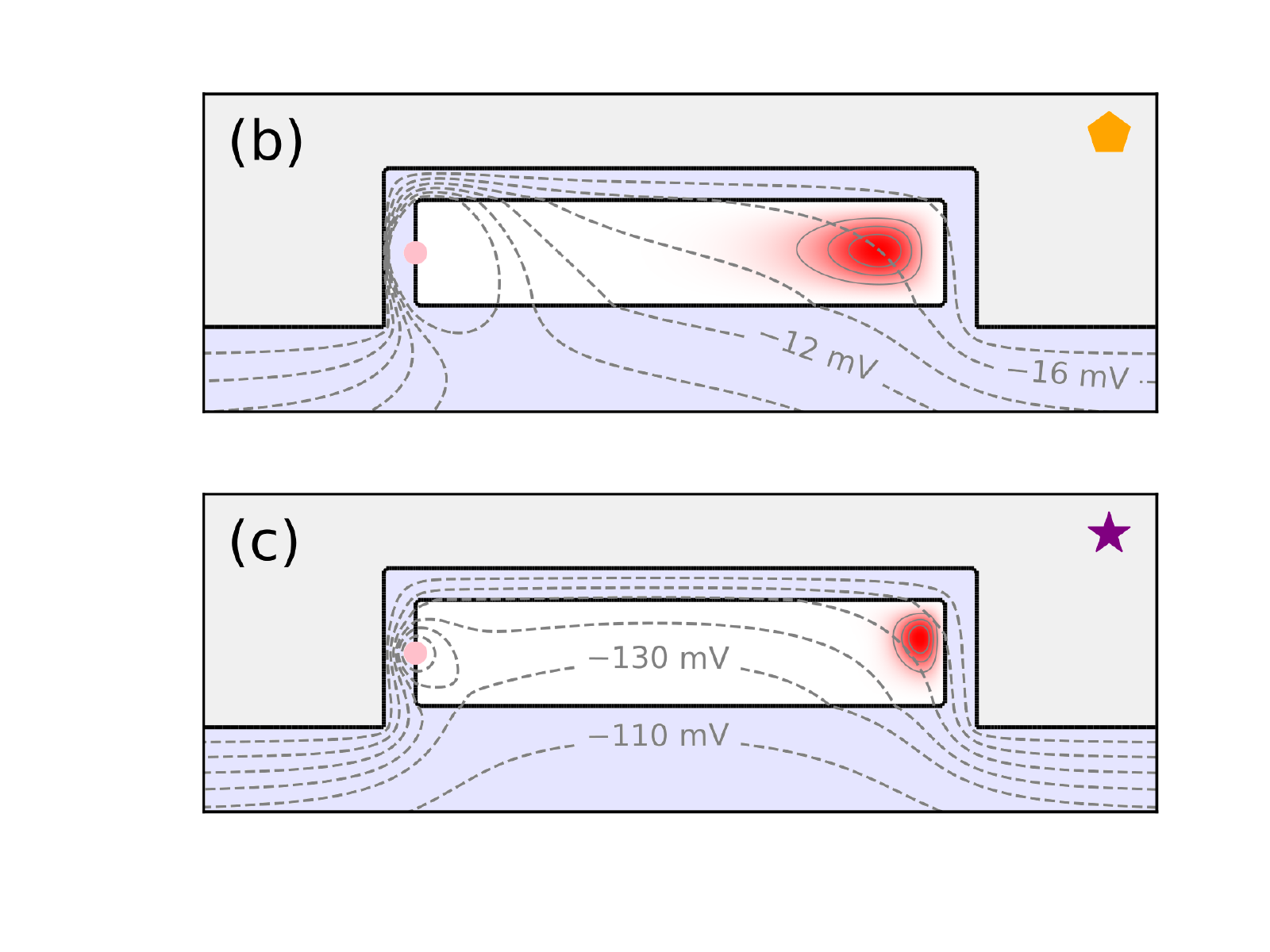}\\
\includegraphics[width = 0.475\textwidth]{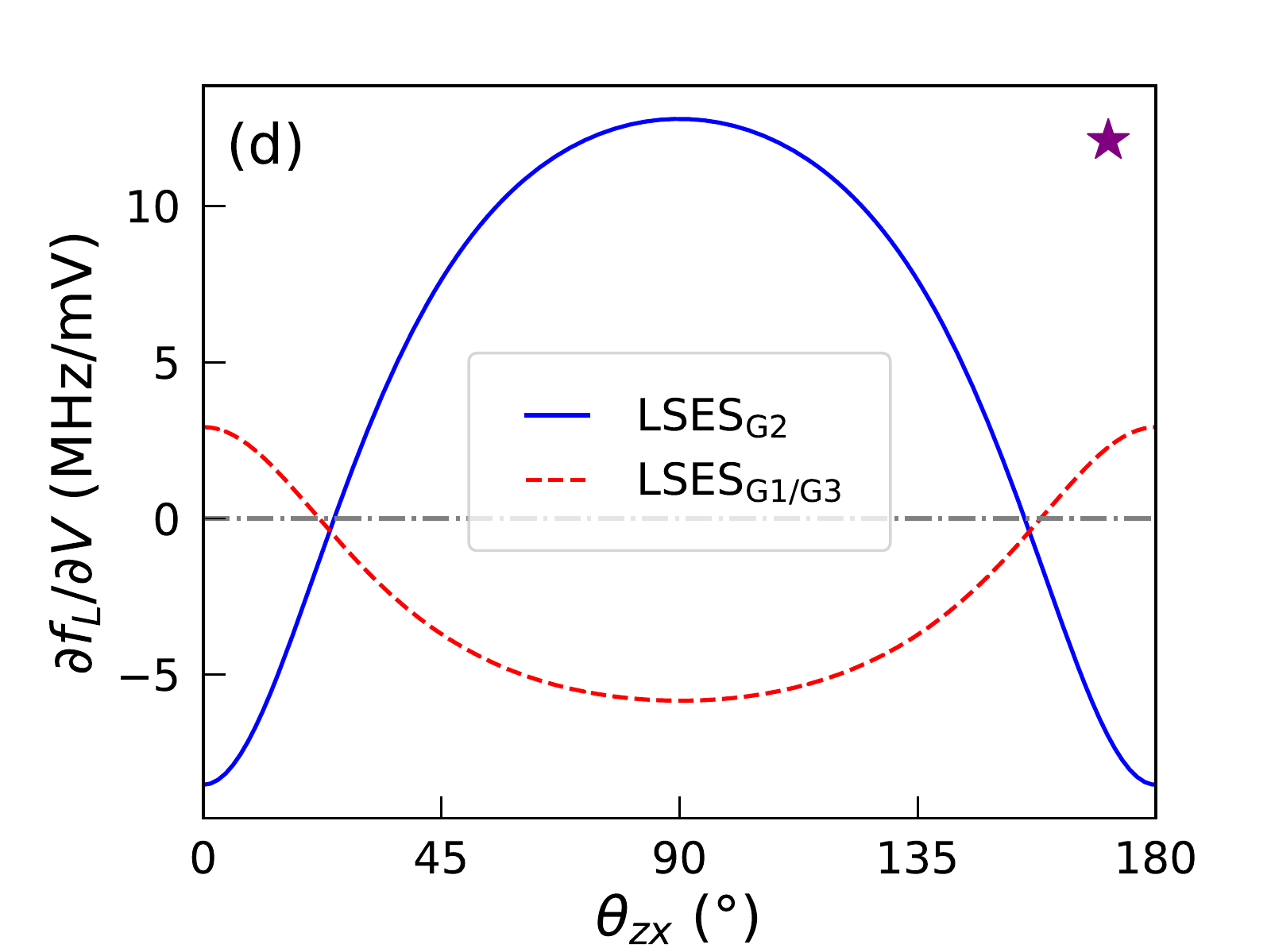}
\caption{\textbf{Dependence of the $g$-factors on the electric field} in a simpler setup with no hole gases below \Gone{}, \Gthree{} and \Gfour{}. \textbf{(a)} $g$-factors $g_x$, $g_y$, and $g_z$ as a function of the difference of potential $-\Vgtwo$ between gates \Gtwo{} and gates \Gone{} and \Gthree{} (both grounded). The larger $-\Vgtwo$, the stronger the vertical and lateral electric fields. \textbf{(b, c)} Maps of the squared wave functions (red) in the cross section of the channel below gate \Gtwo{}, at the biases marked with an orange pentagon and a purple star in (a). The channel is colored in white, the gate \Gtwo{} in gray and SiO$_2$ in blue. The dashed gray lines are isopotential lines of $V_{\rm QD}(\mathbf{r})$, spaced by 2 mV in (b) and by 10 mV in (c). The isodensity surface of the wave function in (c) that encloses $85\%$ of the hole charge is represented in Fig. 1c of the main text. \textbf{(d)} LSES computed at the purple star in (a) as a function of $\theta_{zx}$ (for constant $f_L=17$ GHz). In these calculations, a single positive charge is introduced on the left facet of the channel [pink dot in (b, c)] to lift the degeneracy between the left and right corner dots.} 
\label{fig:gvsVg}
\end{figure*}

In our nanowire, non-planar geometry where each gate covers three facets of the nanowire, there is a significant in-plane electric field component pushing the hole against the lateral $\{1\overline{1}0\}$ facets. This gives rise in principle to two symmetric ``left'' and ``right'' dots hybridized by tunneling across the channel. The calculated tunneling gap in such a large nanowire is, however, below 1 $\mu$eV in the present bias conditions. Therefore, any disorder that splits the left and right sides of the channel by more than a few $\mu$eV leads to the formation of two independent and non-degenerate ``corner'' dots with similar properties \cite{Voisin_NL_2014}. This is illustrated in Fig. \ref{fig:gvsVg}, in a simpler setup with no hole gases under \Gone{}, \Gthree{} and \Gfour{}. There we have added a positive charge on the left facet, which raises the energy of the left dot and break the degeneracy with the right dot; the calculated $g$-factors of the right dot are, however, little dependent on the exact position of the charge introduced on the left side of the channel. Note that there are no clear signatures of a second corner dot in the experimental data, probably because tunneling in and out of this dot occurs at an undetectable rate. As a matter of fact, tunnel rates can be highly sensitive to small perturbations of the potential landscape. Also note that it is practically impossible to determine whether the experimentally observed corner dot is actually on the left or on the right.

Given the width of the device, the hole is very responsive to the lateral electric field, and gets readily squeezed near one of the top corners of the channel, in a dot with comparable vertical and lateral extensions (Fig. \ref{fig:gvsVg}c). The enhancement of lateral with respect to vertical confinement admixes a light-hole envelope into the hole wave function, which results in a decrease of $g_x$ ($\partial g_x/\partial \Vgtwo>0$) and an increase of $g_y$ and $g_z$ ($\partial g_z/\partial \Vgtwo<0$, see Fig. \ref{fig:gvsVg}a) \cite{Venitucci_PRB_2019,Michal_PRB_2021}. The mixing is particularly strong here because the structural vertical confinement is weak ($H=17$ nm) so that the heavy-hole/light-hole gap is small. The $g$-factors (especially $g_x$ and $g_y$) tend to saturate rapidly with increasingly negative $\Vgtwo$ as the heavily squeezed hole hardly responds any more to the vertical and lateral electric fields ($|\partial g_x/\partial \Vgtwo|\ll|\partial g_z/\partial \Vgtwo|$). The LSES computed at the purple star of Fig. \ref{fig:gvsVg}a are plotted as a function of $\theta_{zx}$ in Fig. \ref{fig:gvsVg}d. Interestingly, the zeros of $\LSEStwo{}$ almost coincide with those of $\LSESone{}$ and $\LSESthree{}$. Indeed, most electric field lines connect gate \Gtwo{} to gates \Gone{} and \Gthree{}, so that the Larmor frequency of the hole is primarily a function of $\Vgtwo-(\Vgone+\Vgthree)/2$, and $\partial f_L/\partial \Vgone\approx \partial f_L/\partial \Vgthree \approx -(\partial f_L/\partial \Vgtwo)/2$.\footnote{We emphasize that variations of $\Vgone{}$ do not only move the dot as a whole along the channel, but deform it on the way, which gives rise to the finite $\LSESone{}$. Only joint, opposite variations $\delta\Vgone{}=-\delta\Vgthree{}$ move the dot as a whole with negligible LSES, at least in the absence of hole gases under \Gone{} and \Gthree{}.}

\begin{figure*}[t]
\includegraphics[width = 0.475\textwidth]{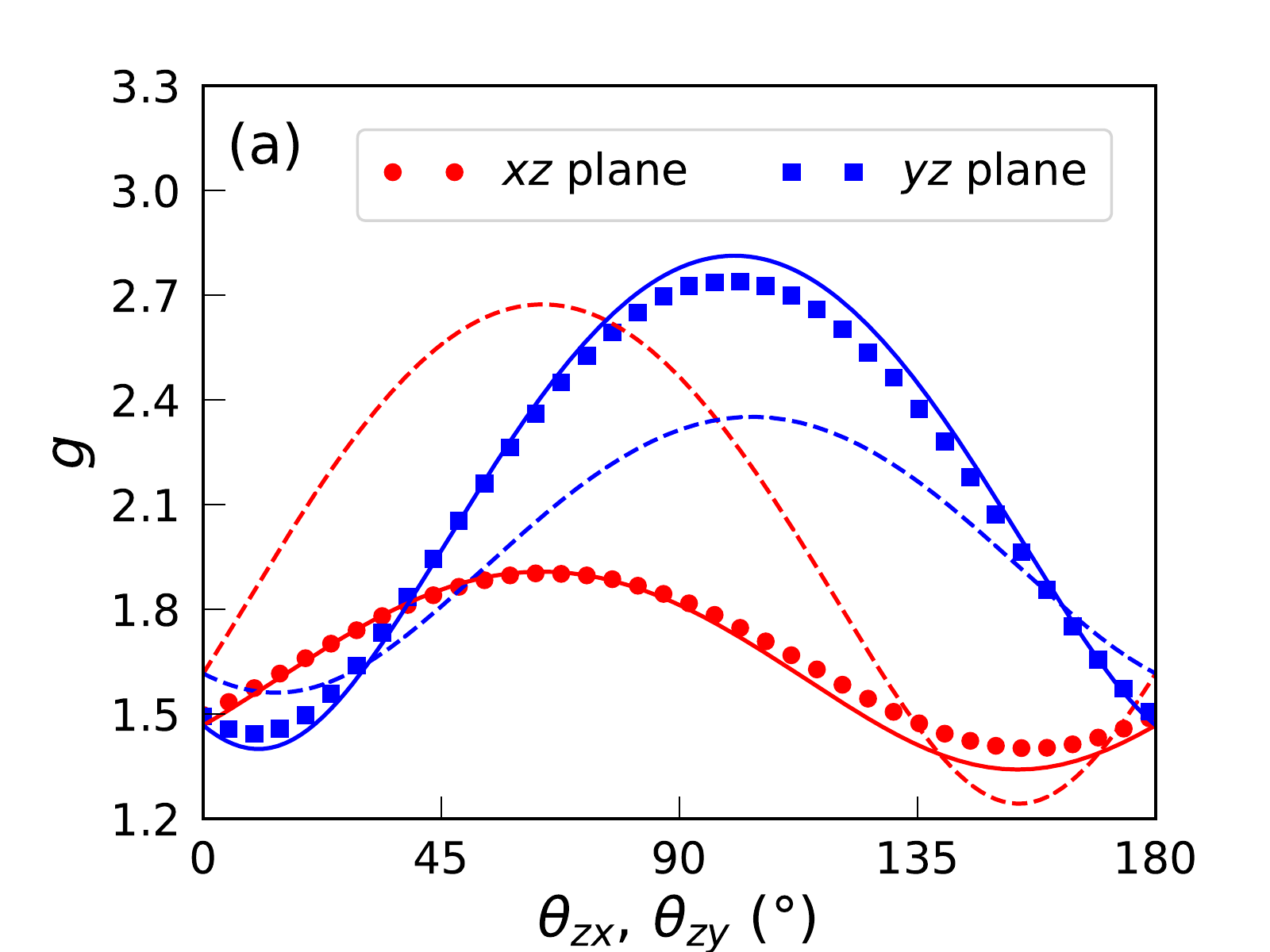}\\
\includegraphics[width = 0.475\textwidth]{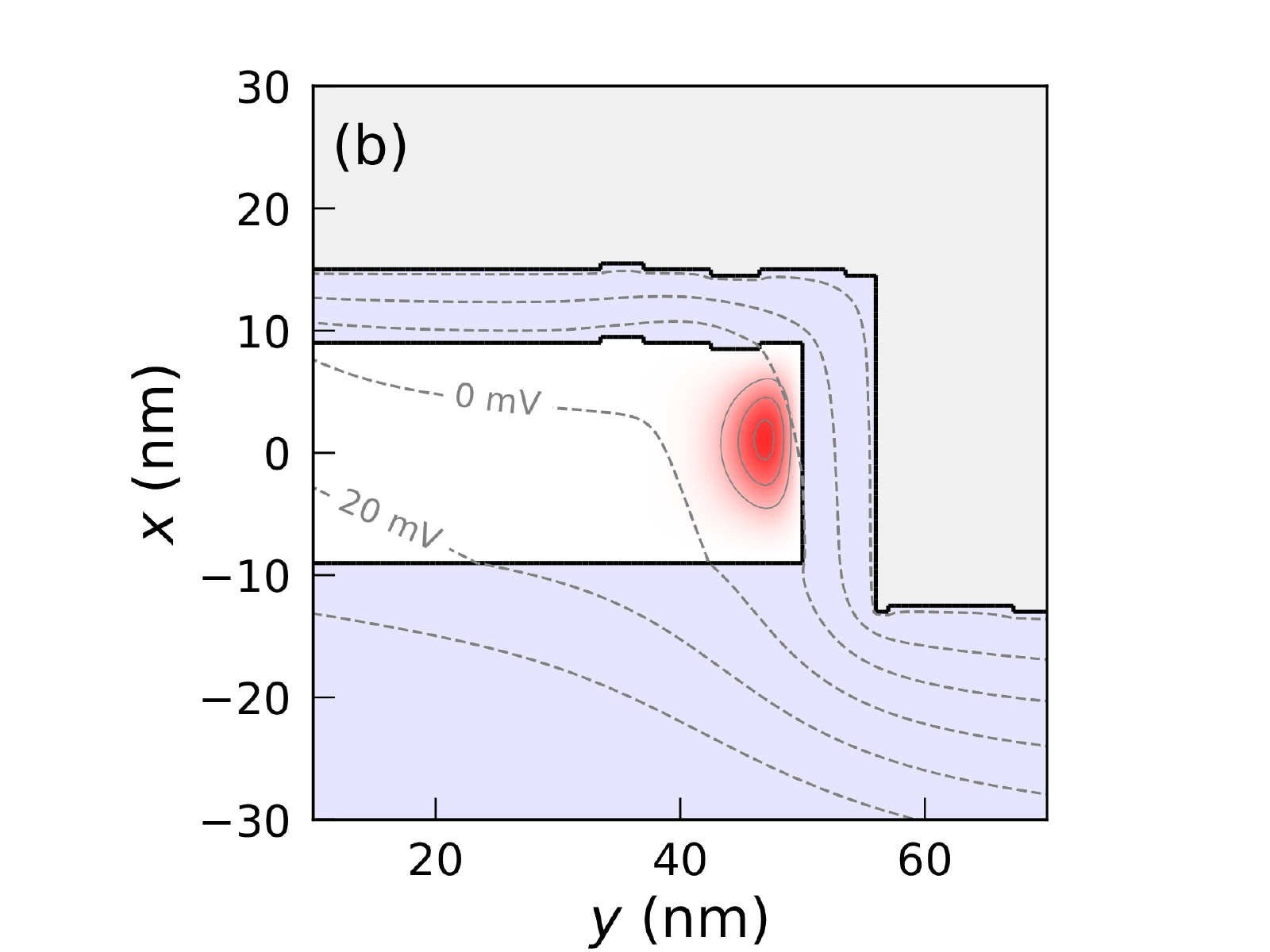}
\includegraphics[width = 0.475\textwidth]{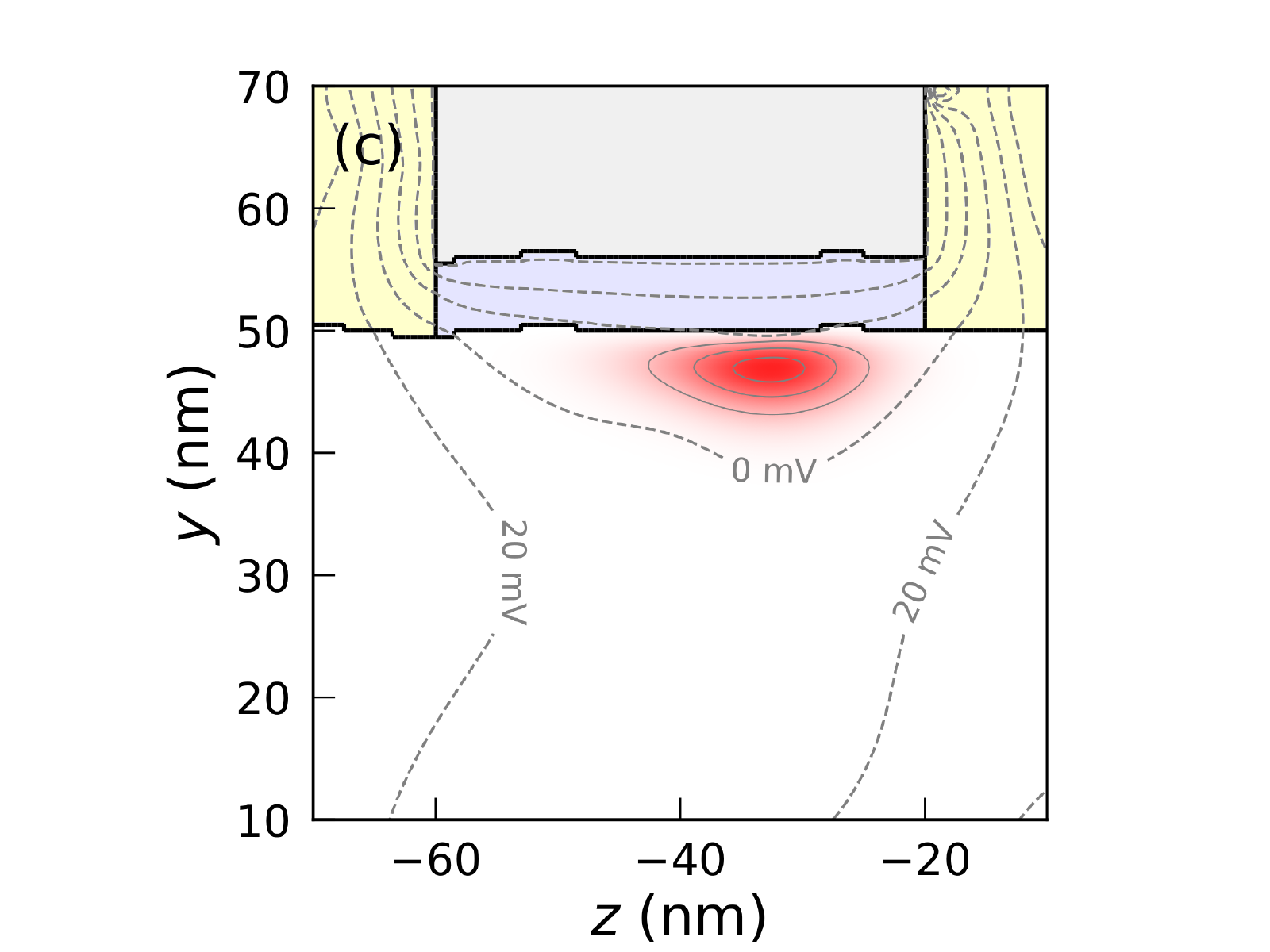}
\caption{\textbf{Comparison between the experimental and calculated $g$-factors.} \textbf{(a)} The $g$-factors are plotted for a magnetic field in the $xz$ (red) and $yz$ (blue) planes, as a function of the angles $\theta_{zx}$ and $\theta_{zy}$, respectively. The symbols are the experimental data; the dotted lines the data calculated in the pristine device; and the solid lines the data calculated in a particular realization of a disordered device with interface roughness and charged traps at the Si/SiO$_2$ interface and in Si$_3$N$_4$ (see text). These traps tend to strengthen confinement on the lateral facets. The polar plots of the $g$-factors and the LSES of this disordered device are shown in Figs. 1 and 2 of the main text, respectively. $\theta_{zx}$ is shifted by $\approx -25\degree$ and $\theta_{zy}$ by $\approx 10\degree$ to account for the experimental rotations of the principal axes of the $g$-tensor (see discussion on strains in section \ref{sec:strains}). \textbf{(b, c)} Maps of the squared wave function (red) computed in the same disordered device, where (b) shows a transverse $xy$ cross section at $z=-35$ nm and (c) a planar $yz$ cross-section at $x=0$. The channel is colored in white, the gate \Gtwo{} in gray, SiO$_2$ in blue and Si$_3$N$_4$ in yellow. The dashed gray lines are isopotential lines of $V_{\rm QD}(\mathbf{r})$, spaced by 20 mV. $V_{\rm QD}(\mathbf{r})$ is here measured with respect to the energy level of the hole.}
\label{fig:comparison}
\end{figure*}

\subsubsection{Enhancement of the lateral electric field by disorder.}

Once screening by the holes gases under \Gone{}, \Gthree{} and \Gfour{} is accounted for, the lateral electric field is too weak to match the measured $g$-factors at the experimental bias point. This is highlighted in Fig. \ref{fig:comparison}a, where the symbols are the experimental $g$-factors and the dashed lines are the calculated ones. $g_y$ remains actually smaller that $g_x$ (at $\theta_{zx}=\theta_{zy}=90^\circ$). This discrepancy may result from inaccuracies in the Thomas-Fermi screening, and (more likely) from additional sources of localization such as disorder. In particular, holes in the channel may be captured by traps at the Si/SiO$_2$ interface ($P_b$ defects) \cite{Biel_2021}, and holes in the poly-silicon gates by traps in the Si$_3$N$_4$ spacers. Such positively charged traps repel the holes and tend to strengthen confinement in the corners, where the resulting potential is best screened by gate \Gtwo{}. The traps are introduced in the simulations as a random distribution of point charges at the Si/SiO$_2$ interface and in Si$_3$N$_4$. We can achieve similar $g$-factors with different combinations of $P_b$ and bulk defects densities; the data displayed in the main text and in Fig. \ref{fig:comparison}a (solid lines) are computed for a particular realization of disorder with density $\sigma_{\rm trap}=5\times 10^{10}$ $P_b$ defects/cm$^2$ at the Si/SiO$_2$ interface and density $\rho_{\rm trap}=5\times 10^{17}$ traps/cm$^3$ in Si$_3$N$_4$. This $\sigma_{\rm trap}$ is typical of Si/SiO$_2$ interface, while the chosen $\rho_{\rm trap}$ does not seem unrealistic given the known affinity of nitrides for charges \cite{Tzeng_JAP_2006}. The potential $V_{\rm QD}(\mathbf{r})$ and the single-hole wave function of this particular device are shown in Figs. \ref{fig:comparison}b,c. The distortions of the isopotential lines and wave function due to disorder are moderate but clearly visible. For the sake of completeness, interface roughness is also included in the simulations. It is characterized by rms fluctuations $\Delta=0.3$ nm and correlation length $L_c=8$ nm \cite{Biel_2021}. The model reproduces the main features of the experimental data, including the magnitude and anisotropy of the $g$-factors and $\LSEStwo{}$ (Fig. 2 of the main text). 

\begin{figure*}[t]
\includegraphics[width = 0.475\textwidth]{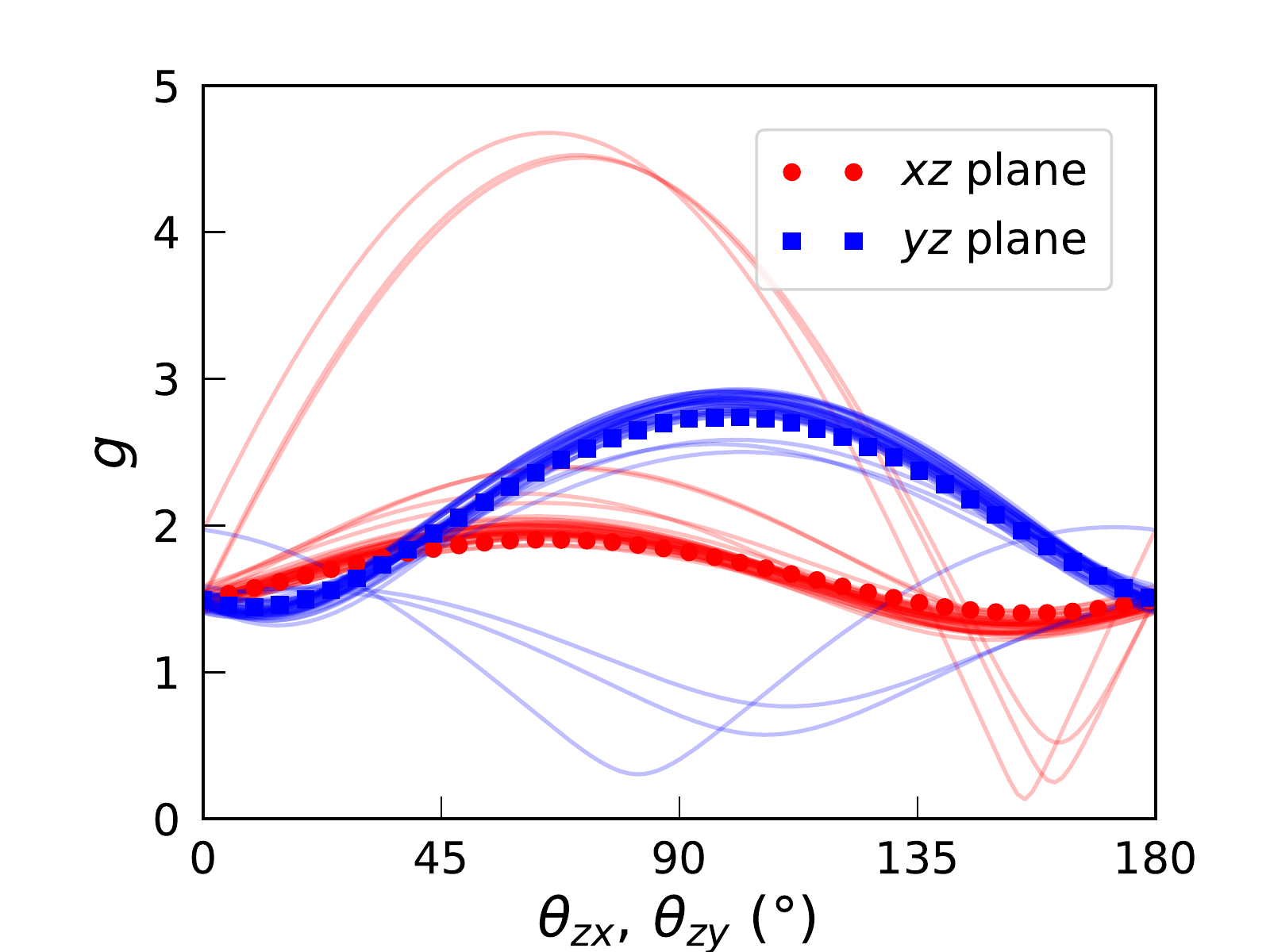}
\caption{\textbf{Variability of the calculated $g$-factors.} Same as Fig. \ref{fig:comparison}; Each line is a different realization of the interface roughness and charge disorder. The interface roughness rms is $\Delta=0.3$ nm and the correlation length is $L_c=8$ nm \cite{Biel_2021}; The density of positively charged traps is $\sigma_{\rm trap}=5\times 10^{10}$ cm$^{-2}$ at the Si/SiO$_2$ interface, and $\rho_{\rm trap}=5\times 10^{17}$ cm$^{-3}$ in Si$_3$N$_4$.}
\label{fig:variability}
\end{figure*}

Choosing $[001]$ as the quantization axis, the hole wave function of Fig. \ref{fig:comparison}b is a strong mixture of heavy ($\approx 54\%\,|3/2,\pm 3/2\rangle_{[001]}$) and light ($\approx 43\%\,|3/2,\pm 1/2\rangle_{[001]}$) envelopes (the reminder being a split-off component). Choosing instead $y=[1\overline{1}0]$ as the quantization axis, the hole appears as a majority $\approx 85\%\,|3/2,\pm 3/2\rangle_{[1\overline{1}0]}$ envelope admixed with a minority $\approx 12\%\,|3/2,\pm 1/2\rangle_{[1\overline{1}0]}$ component. The measured and computed $g_y>g_x$ is the salient fingerprint of the prevalence of $|3/2,\pm 3/2\rangle_{[1\overline{1}0]}$ over $|3/2,\pm 3/2\rangle_{[001]}$ components. The confinement being comparable along $x$ and $y$, the hole actually appears purest when quantized along $z=[110]$, where it stands as a $\approx 90\%\,|3/2,\pm 1/2\rangle_{[110]}$ envelope (as expected from $g_z<g_x,g_y$) \cite{Kloeffel_PRB_2018}.

The disorder gives rise to variability in the $g$-factors (dependence on the particular realization of the disorder \cite{Biel_2021}). This is outlined in Fig. \ref{fig:variability}, which shows the $g$-factors calculated in 50 devices with different samples of disorder. Fourty-four out of the 50 devices still show $g$-factors in reasonable agreement with the experiment. Indeed, the $g$-factors tend to saturate once the hole is squeezed on a lateral facet as discussed above. In the 3 disorder configurations featuring large $g_x$ and small $g_y$, the hole remains localized at the top interface because there are $P_b$ defects near both corners.

Although we can reach a satisfactory agreement with the experimental $\LSEStwo{}$ for many realizations of the disorder, we systematically miss the anisotropy of $\LSESone{}$ (Fig. 2 of the Main Text). This discrepancy may result from limitations of our model. While we capture semi-quantitatively the strong screening of the electric field of gate G1 by the hole gas beneath ($|\partial f_L/\partial \Vgone|\ll|\partial f_L/\partial \Vgtwo|/2$, contrary to Fig. \ref{fig:gvsVg}d), the Thomas-Fermi approximation used to model this hole gas may not be accurate enough. It certainly misses quantization effects as well as the magnetic response of the hole gas. Given the large number of holes under gates G1/G3/G4, going beyond the Thomas-Fermi approximation is however far from trivial. Also, strain inhomogeneities when the dot is moved along the channel may play a role in $\LSESone{}$ (see below). The fact that $\LSESone{}$ is always negative can be explained by the presence of a charged $P_b$ defect in the vicinity of gate G2 that pushes the dot towards gate G3, as shown in Fig. \ref{fig:comparison}c. The robustness of the LSES with respect to disorder will be further discussed in section \ref{sec:outlook}.


\subsubsection{Strains and the rotations of the principal magnetic axes.}
\label{sec:strains}

We finally discuss the misalignment of the principal axes of the $g$-tensor with respect to the device axes. Indeed, the principal axes $X$, $Y$, $Z$ of the calculated $g$-tensor are almost perfectly aligned with the device $x$, $y$ and $z$ axes, whereas those of the experimental $g$-tensor are slightly rotated [by $\approx 10^\circ$ around $x$ ($xyz\to xYz'$), then $\approx -25^\circ$ around $Y$ ($xYz'\to XYZ$)]. The large rotation around $Y$ can hardly be accounted for by a simple misalignment of the sample. The fact that $Z$ is not oriented along the channel implies a loss of the $xy$ quasi-symmetry plane of gate \Gtwo{} \cite{Venitucci_PRB_2018}, and the existence of additional heavy-hole/light-hole mixing mechanisms. The disorder introduced in the previous section actually rotates the principal axes of the $g$-tensor, but by no more than a few degrees. The most likely scenario is that \Qtwo{} is slightly displaced towards \Gthree{} (as suggested above), and experiences small process and cool-down strains \cite{Venitucci_PRB_2018,liles_2020}. In particular, shear strains control the phase of the heavy-hole/light-hole mixing matrix elements. In the basis set and axes set of Ref. \citenum{Michal_PRB_2021}, they give rise to non-diagonal corrections to the $g$-matrix:
\begin{equation}
\delta g_{zy}\approx\frac{4\sqrt{3}\kappa d}{\Delta}\varepsilon_{yz},\,\delta g_{zx}\approx\frac{4\sqrt{3}\kappa d}{\Delta}\varepsilon_{xz},\,\delta g_{xy}=-\delta g_{yx}\approx -\frac{12\kappa b}{\Delta}\varepsilon_{xy}\,,
\end{equation}
where $b=-2.1$ eV and $d=-4.85$ eV are the uniaxial and shear deformation potentials of the valence band of silicon, and $\Delta$ is the heavy-hole/light-hole gap. Therefore, the shear strains $\varepsilon_{yz}$, $\varepsilon_{xz}$, and $\varepsilon_{xy}$ drive rotations of the principal magnetic axes around $x$, $y$ and $z$ respectively. Our simulations reproduce the experimental rotations assuming small $\varepsilon_{yz}\simeq 0.03\%$ and $\varepsilon_{xz}\simeq 0.08\%$, which highlights the sensitivity of such quantum devices to residual strains \cite{Venitucci_PRB_2018, Pla_PRApp_2018,liles_2020}. The assessment of the inherently inhomogeneous strains in such complex nanostructures remains, however, difficult (in particular in the nitrides), and goes beyond the scope of this work. In the absence of a complete distribution of strains (including hydrostatic and uniaxial components), we have practically shifted $\theta_{zx}$ by $\approx -25\degree$ and $\theta_{zy}$ by $\approx 10\degree$ in the simulations of Figs. 1 and 2 of the main text, and in Figs. \ref{fig:comparison} and \ref{fig:variability}, as if these rotations resulted from a misalignment of the sample with respect to the magnet axes. Note that the possible rotation of the principal axes around $z$ can not be resolved since the $g$-factors have not been measured in the $xy$ plane; yet it must be within $\pm 20\degree$ to reach a satisfactory agreement between theory and experiment.

Since uniaxial and shear strains rule the heavy-hole/light-hole mixing together with confinement, they can in principle help reduce the lateral confinement, hence the disorder needed to reach agreement with the experimental $g$-factors. We emphasize, though, that the dot becomes much more responsive to \Gtwo{} once deconfined from the corner, so that $\partial f_L/\partial \Vgtwo$ increases significantly. Therefore, the magnitude of the experimental $\partial f_L/\partial \Vgtwo$, as well as the fact that the experimental $g$-factors match the saturation values calculated at large gate voltage/electric field (Fig. \ref{fig:gvsVg}), support strong confinement in the corners and small strains. 

To conclude, the present model captures and explains the most salient features of the experiment: the anisotropy of the $g$-factors, $g_y>g_x>g_z$, and of $\partial f_L/\partial \Vgtwo$ result from the balance between vertical and lateral confinement in the corner dot of a ``thick'' silicon film; $\partial f_L/\partial \Vgone$ is strongly screened by the hole gas accumulated under gate \Gone{} and is, therefore, much smaller (in magnitude) than $\partial f_L/\partial \Vgtwo$. The remaining discrepancies (in particular the rotation of the principal axes of the $g$-tensor) are attributed to residual process and cool-down strains and to possible inaccuracies in the description of screening.

\begin{figure*}[t]
\includegraphics[width = 0.475\textwidth]{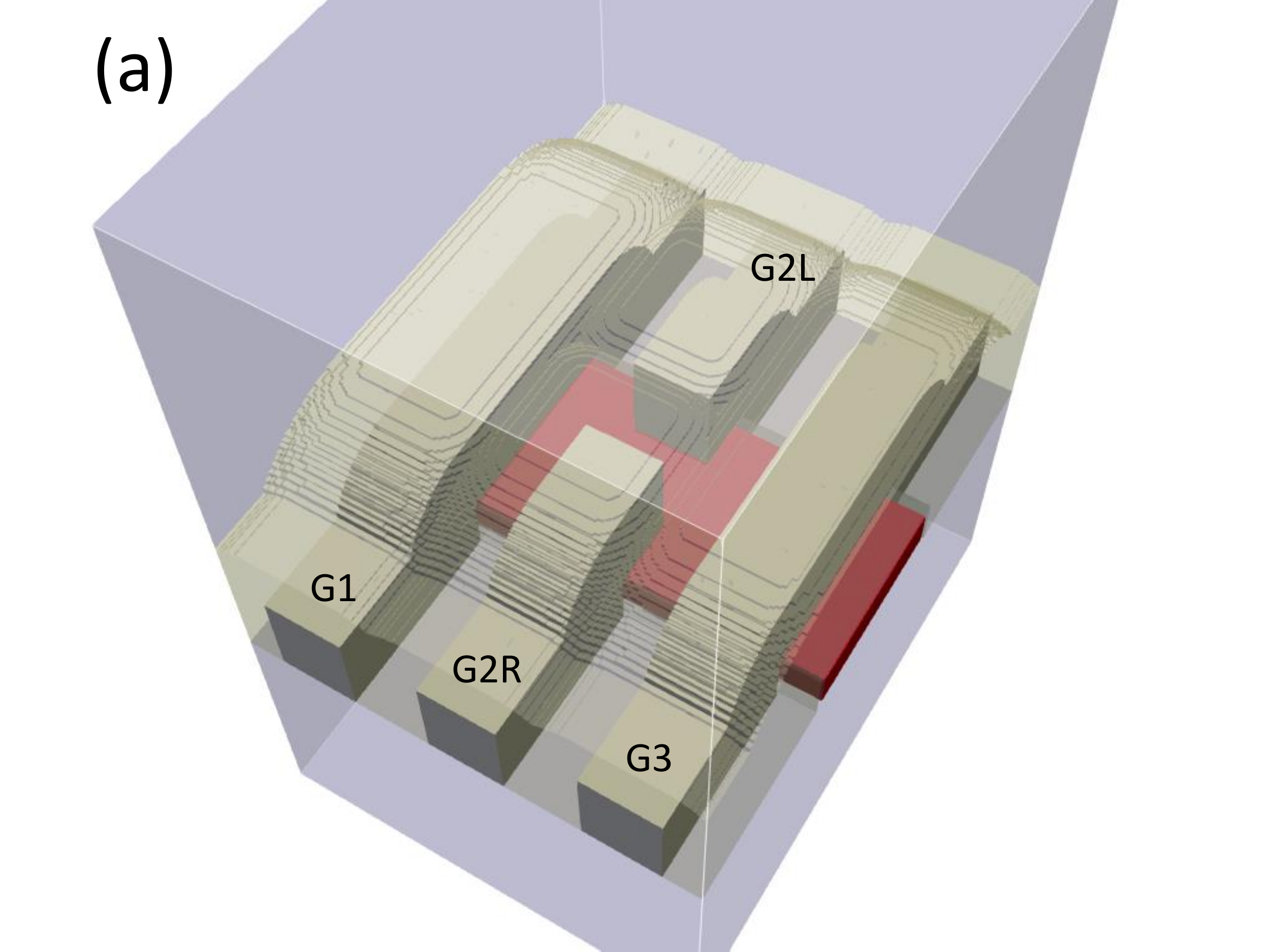} \\
\includegraphics[width = 0.475\textwidth]{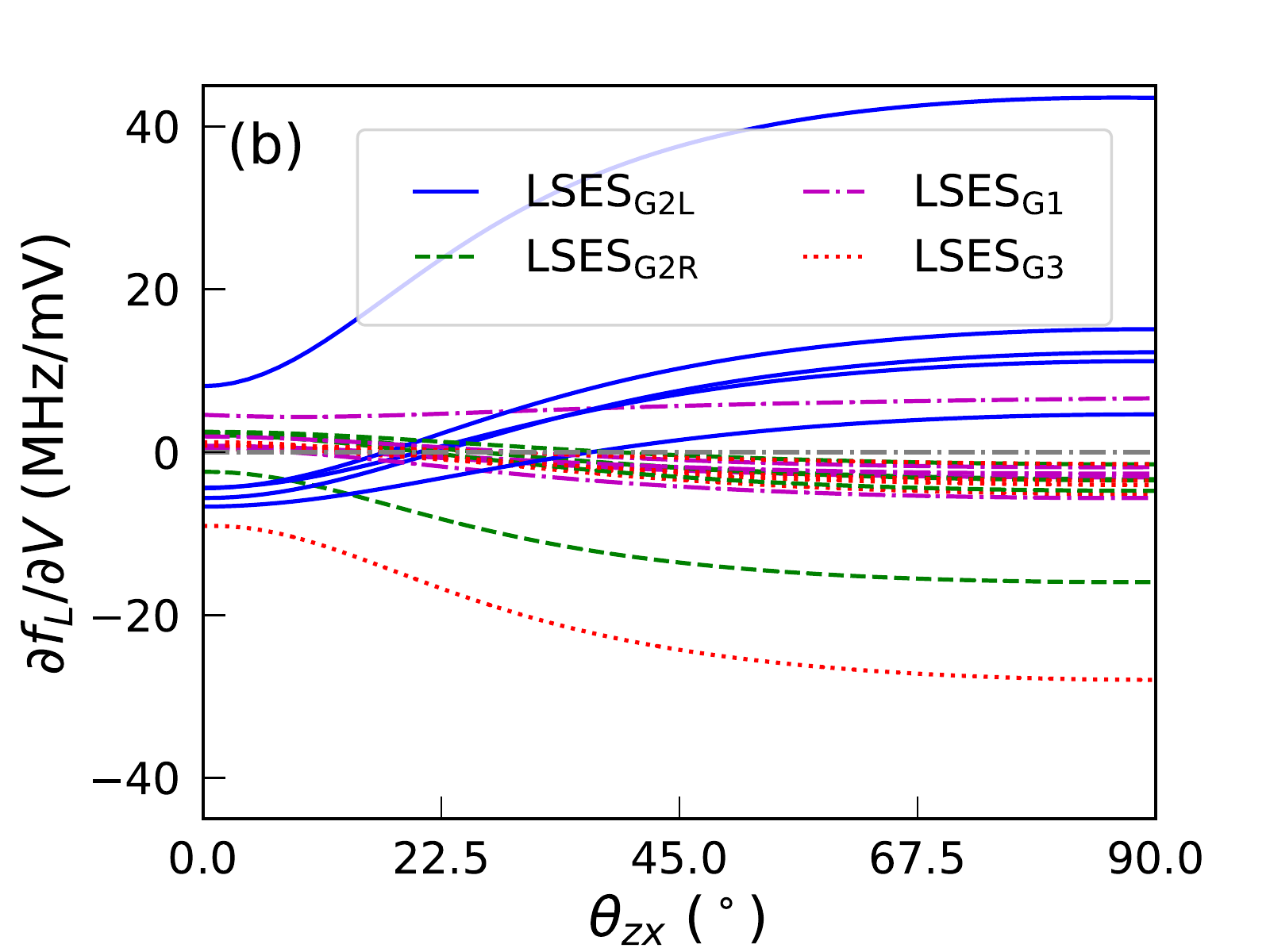}
\includegraphics[width = 0.475\textwidth]{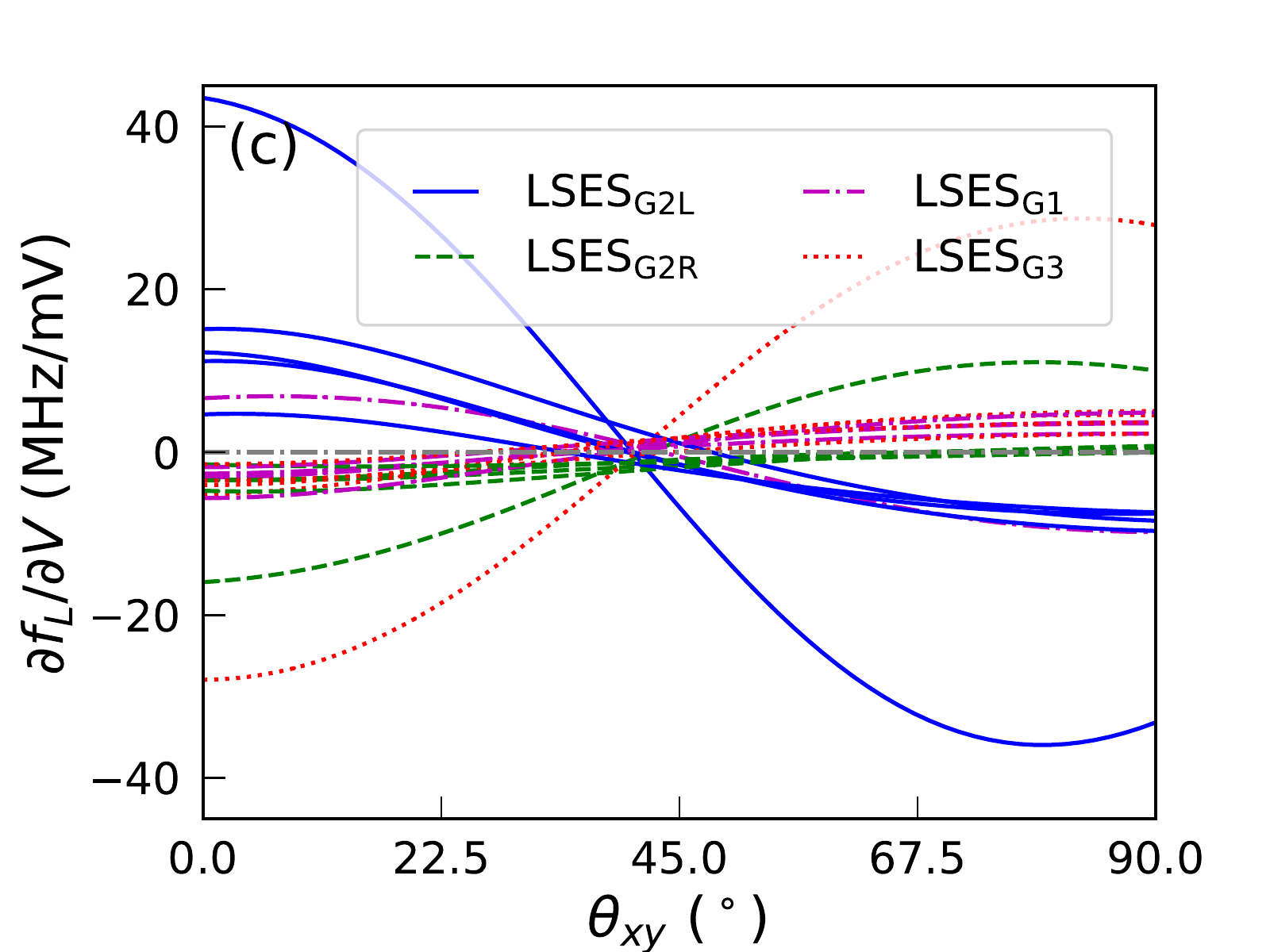}

\caption{\textbf{LSES of a face-to-face device.} \textbf{(a)} A face-to-face device similar to Fig. \ref{fig:DeviceModel}, but with G2 split into a left (G2L) and a right (G2R) gate separated by 40 nm. \textbf{(b)} LSES$_{\rm G1}$, LSES$_{\rm G2L}$, LSES$_{\rm G2R}$ and LSES$_{\rm G3}$ as a function of the angle $\theta_{zx}$ between the $z$ axis and the magnetic field $\mathbf{B}$ in the $xz$ plane, for five different random distributions of charge traps at the Si/SiO$_2$ interface ($\sigma_{\rm trap}=5\times 10^{10}$ cm$^{-2}$). The Larmor frequency is set to $f_L=10$ GHz. \textbf{(c)} Same as (b) as a function of the angle $\theta_{xy}$ between the $x$ axis and the magnetic field $\mathbf{B}$ in the $xy$ plane.}
\label{fig:deviceF2F}
\end{figure*}

\subsection{Outlook: Robustness of the sweet spots with respect to disorder}
\label{sec:outlook}

As an outlook, we investigate the robustness of the sweet spots evidenced in this work. For that purpose, we consider a more versatile ``face-to-face'' layout \cite{Roche_PRL_2012} where \Gtwo{} is split into two independent gates $G_{\rm 2L}$ and $G_{\rm 2R}$ that overlap the left (L) and right (R) corners respectively (see Fig. \ref{fig:deviceF2F}a). In this configuration, the corner dots are much more deterministic and stable since the potential on the left and right sides of the channel can be adjusted independently. To probe the robustness of possible sweet spots, we introduce positive charge traps at the surface of silicon with density $\sigma_{\rm trap}=5\times 10^{10}$ cm$^{-2}$.

We ground all gates except G2L ($V_{\rm G2L}=-50$ mV). In such a thick channel ($H=20$ nm), the lateral electric field between gates G2R and G2L is already large enough to squeeze the dot on the lateral facet and reach the regime $g_y>g_x$ (alternatively, the back gate voltage can be made positive to strengthen confinement in the corners). We next compute the LSES with respect to the four gates (LSES$_{\rm G1}$, LSES$_{\rm G2L}$, LSES$_{\rm G2R}$ and LSES$_{\rm G3}$). The data collected for five representative configurations of disorder are plotted in Fig. \ref{fig:deviceF2F}b,c for a magnetic field $\mathbf{B}$ in the $xz$ and $xy$ planes, respectively. In Fig. \ref{fig:deviceF2F}c, $\theta_{xy}$ is the angle between the magnetic field $\mathbf{B}$ and the $x$ axis. The two planes exhibit qualitative differences with respect to the variability of the LSES. In particular, the sweet spot in the $xz$ plane is generally more sensitive to disorder (and even missing for one of the configurations). On the contrary, there is a remarkably robust sweet spot in the $xy$ plane near $\theta_{xy}\approx 41\degree$.

The sweet spots in the $xz$ and $xy$ planes actually belong to the same ``sweet line'' running around the $x$ axis \cite{Michal2022}. The sweet spot is however more robust to disorder in the $xy$ plane because in this class of devices $\partial g_x/\partial V\approx -\partial g_y/\partial V$ whatever the gate\footnote{In the language of Ref. \citenum{Michal2022}, $\left|\frac{\partial\gt}{\partial V}\cdot\mathbf{b}\right|$ is weakly dependent on the orientation $\mathbf{b}=(\cos\theta_{xy},\,\sin\theta_{xy},\,0)$ of the magnetic field in the $(xy)$ plane, with $\gt$ the $g$-matrix.} -- in other words, the in-plane electric field primarily shifts weight between $g_x$ and $g_y$ (also see Fig. \ref{fig:gvsVg}a). Therefore, for a given $\theta_{xy}$, the derivative of the $g$-factor with respect to the gate voltage reads
\begin{equation}
\frac{\partial g}{\partial V}=\frac{\partial}{\partial V}\sqrt{g_x^2\cos^2\theta_{xy}+g_y^2\sin^2\theta_{xy}}=\frac{1}{g}\left(g_x\frac{\partial g_x}{\partial V}\cos^2\theta_{xy}+g_y\frac{\partial g_y}{\partial V}\sin^2\theta_{xy}\right)\approx \left(g_x\cos^2\theta_{xy}-g_y\sin^2\theta_{xy}\right)\frac{1}{g}\frac{\partial g_x}{\partial V}\,,
\end{equation}
which is zero when $\theta_{xy}\approx\frac{\pi}{2}\pm\arctan\sqrt{\frac{g_y}{g_x}}$. Hence, the position of the sweet spot is resilient to moderate disorder once $g_x$ and $g_y$ get close to saturation. Despite this saturation, the Rabi frequencies of the device of Fig. \ref{fig:deviceF2F} are still in the $10$ MHz range for a 1 mV drive on gate G2L. They are actually maximal near the sweet spot in the $xy$ plane \cite{Michal2022} even in the presence of disorder.\footnote{The sweet spot in the $xy$ plane could not be probed with the present experimental setup, see Supp. Info \ref{suppinf:Setup}.}

We expect such robust sweet spots to be ubiquitous in a large variety of silicon and germanium devices \cite{Michal2022,Bosco_PRXQ_2021}. Therefore, device optimization and improvements in material quality shall further enhance the performances of hole spin qubits in the near term.

\section{\label{suppinf:Spin electric susceptibility} LSES with respect to gate 2 ($\LSEStwo$)}

\begin{figure*}[h]
\includegraphics[width = 0.98\textwidth]{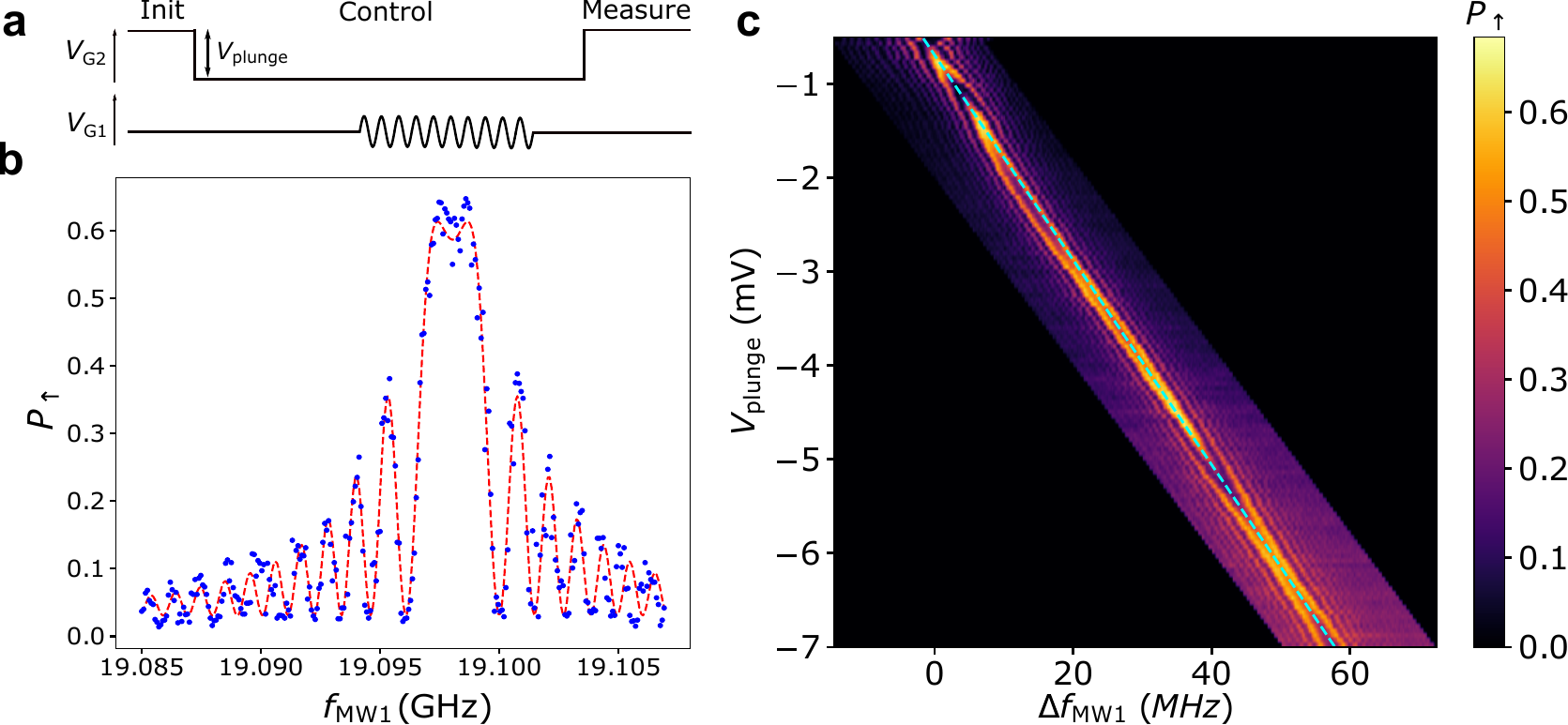}
\caption{\label{fig:LSES2} \textbf{Measurement of $\LSEStwo$.} \textbf{(a)} Schematic representation of the pulse sequence used to monitor spin resonance. We burst on MW1 for $5$\,$\mu$s and average $P_{\uparrow}$ over 200 such sequences. \textbf{(b)} Average $P_{\uparrow}$ (Blue dots) versus MW1 burst frequency at $V_{\rm plunge}=-1$\,mV. This plot is in essence a line cut of a Rabi chevron at $t_{\rm burst}=5$\,$\mu$s. The red dashed line is a fit used to extract the Larmor frequency. \textbf{(c)} Tracking of $f_L$ as a function of $V_{\rm plunge}$. The dashed blue line is a linear fit whose slope is equal to $\LSEStwo$.}
\end{figure*}

To measure $\LSEStwo$, we apply a two stage sequence (Initialisation/Measure and Control) on MW2 while bursting for $5$\,$\mu$s on MW1 to drive coherent spin rotations (see Fig.~\ref{fig:LSES2}a). We record the oscillations of $P_{\uparrow}$ (averaged over 200 pulse sequences) as a function of the MW1 burst frequency $f_{\rm MW1}$ (Fig.~\ref{fig:LSES2}b), and fit with a Rabi chevron model to extract the Larmor frequency $f_L$. 

We repeat the experiment for different $V_{\rm plunge}$, and obtain the map of Fig.~\ref{fig:LSES2}c, where $\LSEStwo=\partial f_L/\partial V_{\rm plunge}$ is the slope of the dashed blue line. Note that the Rabi frequency also depends on $V_{\rm plunge}$.

\section{\label{suppinf:Rabi} Rabi oscillations at the sweet spot}

\begin{figure*}[h]
\includegraphics[width = 1\textwidth]{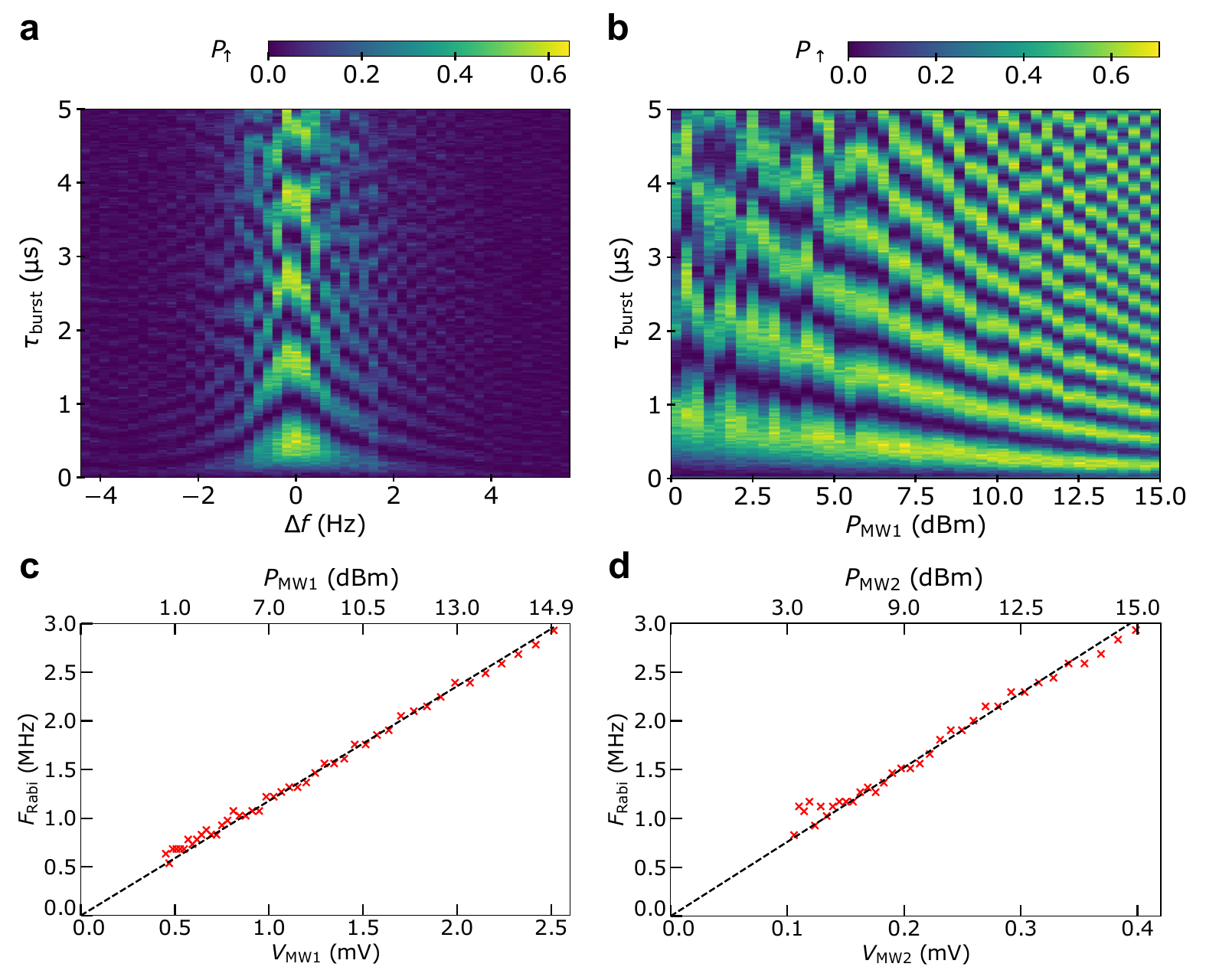}
\caption{\label{fig:rabi} \textbf{Electrical spin driving at the coherence sweet spot} \textbf{(a)} Chevron pattern at $\theta_{zx}=99\degree$ recorded with the same pulse sequence as in Fig.~\ref{fig:LSES2}a. $P_{\uparrow}$ is plotted versus MW1 detuning from spin resonance ($\Delta f=f_{\rm MW1}-f_L$) and MW1 burst duration $\tau_{\rm burst}$. The Larmor frequency is $f_L=17$\,GHz and the MW1 power on top of the fridge is $P_{\rm MW1}=5$\,dBm. \textbf{(b)} $P_{\uparrow}$ versus $P_{\rm MW1}$ and $\tau_{\rm burst}$ for $\Delta f=0$. \textbf{(c)} Rabi frequency extracted from (b) versus on chip MW1 amplitude $V_{\rm MW1}$ (symbols) assuming 30 dB attenuation from attenuators and 30 dB loss from cables at low temperature. The top axis is the power $P_{\rm MW1}$ delivered on top of the fridge. The dashed line is a linear fit with slope $F_{\rm Rabi}=1.2$\,MHz/mV, which evidences the absence of saturation at least up to 3\,MHz. \textbf{(d)} Same as (c) but driving the spin using \Gtwo. The attenuation is larger on this line (46\,dBm), so that the Rabi oscillations are actually 6 times faster on \Gtwo{} (slope $F_{\rm Rabi}=7.6$\,MHz/mV) than on \Gone{}.}
\end{figure*}

Sweet spots for coherence may result from a simple zero of the longitudinal spin-electric susceptibility, or from the complete decoupling of the hole from the electric field (for example if the wave function becomes centrosymmetric \cite{Venitucci_PRB_2018,Michal2022}). In the latter case, Rabi oscillations (transverse spin-electric susceptibility) are also impossible. Figure~\ref{fig:rabi} demonstrates that the hole can still be manipulated electrically near the sweet spot for coherence at $\theta_{zx}=99\degree$. In the experiment reported in the main text, the hole is driven by a microwave burst on gate \Gone{}. The Rabi frequency is found dependent on the magnetic field orientation (see Fig. \ref{fig:quality_factor}a of section \ref{suppinf:Qfactor}), with a minimum around the sweet spot, where the hole spin still rotates up to $F_{\rm Rabi}=5$ MHz for an applied power of 20 dBm on top of the MW1 line. A microwave burst on gate \Gtwo{} also enables spin rotation up to $\sim$ 3 MHz at the sweet spot (see Fig.~\ref{fig:rabi}d). However, we speculate that the Rabi frequency is only limited by the available microwave power and the line attenuation, since we do not observe any saturation with increasing power. After conversion of the microwave power into gate voltage amplitudes, we find that the driving efficiency is much larger on gate \Gtwo{} ($F_{\rm Rabi}=7.6$ MHz/mV) than on gate \Gone{} ($F_{\rm Rabi}=1.2$ MHz/mV), which suggests that the spin could be rotated much faster by reducing the attenuation on the MW2 line.

\vspace{2cm}
\section{\label{suppinf:Qfactor}Quality factors} 

\begin{figure*}[h!]
\includegraphics[width=0.95 \textwidth]{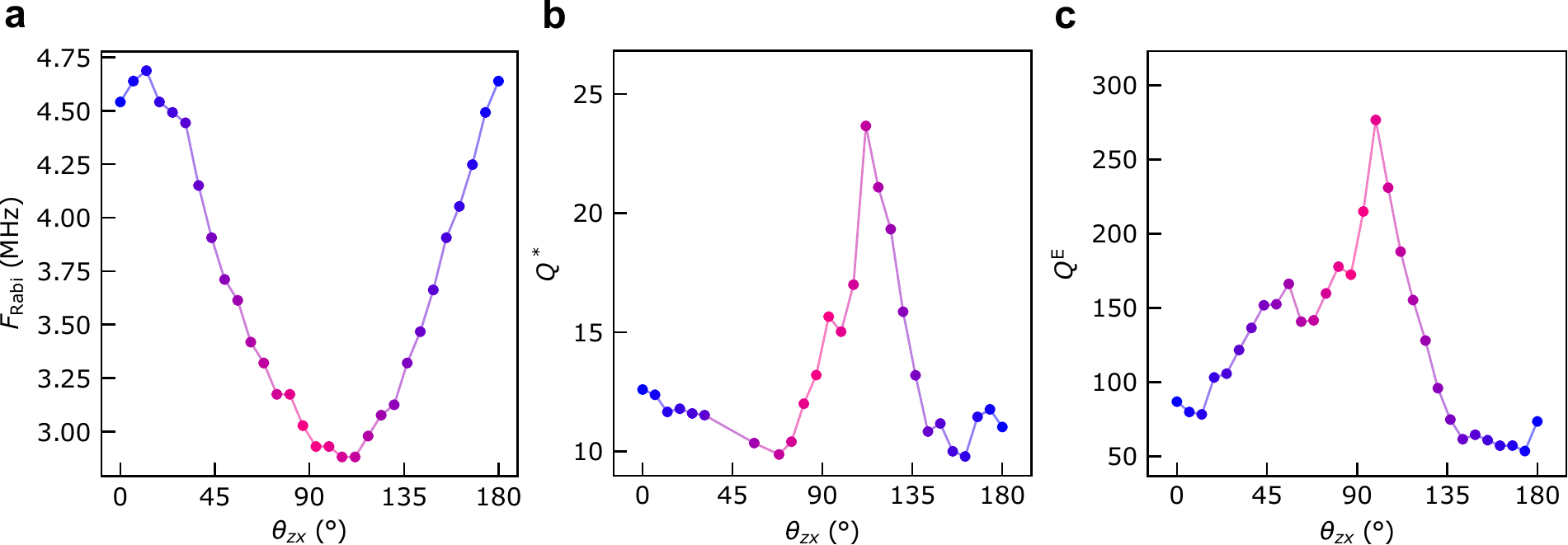}
\caption{\label{fig:quality_factor} 
\textbf{Rabi frequencies and quality factors.} \textbf{(a)} Rabi frequency as a function of magnetic field orientation $\theta_{zx}$. The Larmor frequency $f_L=17$ GHz is kept constant and the hole spin is manipulated by a microwave burst on gate \Gone{} with power $P_{\rm MW1}=15$ dBm on top of the MW1 line. \textbf{(b)} Inhomogeneous quality factor $Q^*$ as a function of the magnetic field orientation $\theta_{zx}$. The data are calculated from the Rabi frequencies plotted in (a), and from the values of $\overline{T}_2^*$ measured for $t_{\rm meas}=5.5$\,s (Fig.~\ref{fig:t2star}b). \textbf{(c)} Same as (b) for the echo quality factor $Q^{\rm E}$.}
\end{figure*}

In this section, we discuss the quality factors of the hole spin. We define:
\begin{itemize}
    \item the inhomogeneous quality factor $Q^*=F_{\rm Rabi}\times T_2^*$, which is half the number of $\pi$ rotations that can be achieved within the inhomogeneous dephasing time $T_2^*$. 
    \item the echo quality factor $Q^{\rm E}=F_{\rm Rabi}\times T_2^{\rm E}$, which is half the number of $\pi$ rotations that can be achieved within the echo time $T_2^{\rm E}$ shall the manipulations be intertwinned with a Hahn-Echo noise decoupling sequence.
\end{itemize}
All quantities involved in the different quality factors depend on the magnetic field orientation. In Figure \ref{fig:quality_factor}a, we plot the Rabi frequency as a function of $\theta_{zx}$ at constant Larmor frequency $f_L=17$ GHz. The spin is driven by microwave bursts on gate \Gone{}, with power $P_{\rm MW1}=15$ dBm (on top of the MW1 line). The resulting quality factors $Q^*$ and $Q^{\rm E}$ are plotted in Figs. \ref{fig:quality_factor}b,c. For $Q^*$, we use the value of $\overline{T}_2^*$ measured at $t_{\rm meas}=5.5$\,s (see section \ref{suppinf:T2nonergodic}). In the present case, the Rabi frequency is minimal around the sweet spot (see \ref{fig:quality_factor}a). Nonetheless, the quality factors $Q^*$ and $Q^{\rm E}$ do peak near the sweet spot owing to the much improved coherence times. They reach $Q^*=23$ and $Q^{\rm E}=276$, with peak-to-valley ratios of respectively $\approx 2.5$ and $\approx 5.5$.

As discussed in section \ref{suppinf:Rabi}, with a larger driving power $P_{\rm MW1}=20$ dBm, we can achieve Rabi frequencies of at least 5 MHz at the sweet spot, which results in $Q^*\approx 35$ and $Q^{\rm E}\approx 440$. In principle, the quality factors may be further improved by driving with gate \Gtwo{} and looking for the sweet spot in the $xy$ plane (see section \ref{sec:outlook}).

\section{\label{suppinf:funny_calculations} Pure dephasing with uncorrelated noise sources}

The hole can generally be described as an effective spin $1/2$ with Hamiltonian \cite{Ares_PRL_2013}
\begin{equation}\label{HS}
H_s=\mathbf{S}\cdot\boldsymbol{\omega}_L(\mathbf{V}_{\rm G})\,.
\end{equation}
Here $\mathbf{S}=\frac{\hbar}{2}\boldsymbol{\sigma}$ is the spin 1/2 operator and $\boldsymbol{\omega}_L(\mathbf{V}_{\rm G})=\frac{\mu_B}{\hbar}\gt(\mathbf{V}_{\rm G})\cdot\mathbf{B}$ stands for the spin precession (Larmor) vector, proportional to the product of the voltage-dependent $g$-tensor (or $g$-matrix \cite{Abragam1970}) $\gt(\mathbf{V}_{\rm G})$ with the external magnetic field $\mathbf{B}$. $\mathbf{V}_{\rm G}=(V_{\rm G1},V_{\rm G2},\dots,V_{{\rm G}n})$ is the set of voltages on gates G1, G2, $\dots$, G$n$. Each can be split into static and dynamical contributions $V_{{\rm G}i}(t)=V_{{\rm G}i}^0+\delta V_{{\rm G}i}(t)$, V$_{{\rm G}i}^0$ being the bias voltage on gate G$i$ and $\delta V_{{\rm G}i}(t)$ the voltage noise responsible for qubit relaxation and decoherence.

The gate voltage noise introduces a random component $\delta\phi(t)$ to the qubit phase $\phi(t)=2\pi f_L t+\delta\phi(t)$, where $f_L=\frac{\mu_B}{h}|\gt(\mathbf{V}_{\rm G}^0)\cdot\mathbf{B}|$ is the Larmor frequency. After free evolution over time $t$ the accumulated random phase reads at first order in the noise \cite{Ithier_PRB_2005, Paladino_Revmodphys_2014}:
\begin{equation}
\delta\phi(t) = 2\pi\int_{0}^{t} dt'\,\delta f_L(t') = 2\pi\int_{0}^{t}dt'\,\sum_i D_{{\rm G}i}\delta V_{{\rm G}i}(t')\,. 
\end{equation}
where $D_{{\rm G}i}=\partial f_L/\partial V_{{\rm G}i}^0$ is the LSES of gate G$i$. More generally, for a dynamical decoupling pulse sequence the accumulated phase is \cite{CP1954, MG1958, Vandersypen2005, Ithier_PRB_2005, Paladino_Revmodphys_2014}:
\begin{equation}
    \delta\phi(t)=2\pi\int_{-\infty}^{+\infty}dt'\,\sum_i D_{{\rm G}i}\delta V_{{\rm G}i}(t')\filter(t')\,,
\end{equation}
where the function $\filter(t')$ describes the effects of the pulse sequence performed over time $t$. In particular, for free induction decay (Ramsey experiment),
\begin{equation}
\filter^{\rm R}(t')=\begin{cases}
1 & \mbox{ if } 0<t'<t, \\
0 & \mbox{ otherwise}\,,
\end{cases}
\end{equation}
and for a CPMG sequence with $N_\pi$ $\pi$-pulses \cite{Cywinski_PRB_2008}:
\begin{equation}
\filter^{\rm CPMG}(t')=\sum_{k=0}^{N_\pi} (-1)^k\theta(t_{k+1}-t')\theta(t'-t_k)\,,
\end{equation}
where $\theta$ is the Heaviside function, $t_k=(k-1/2)t/N_\pi$ for $k=1,\dots,N_\pi$, and by definition \cite{Cywinski_PRB_2008} $t_0=0$ and $t_{N_\pi+1}=t$. The Ramsey and the Hahn echo experiments are particular cases of the CPMG sequence with $N_\pi=0$ and $N_\pi=1$ respectively.

The dephasing experienced by the spin as a consequence of voltage noise is characterized by the decay of the off-diagonal element of the spin density matrix in the rotating frame \cite{Paladino_Revmodphys_2014}:
\begin{equation}
\langle\tilde{\rho}_{01}\rangle(t)=\tilde{\rho}_{01}(0)\langle e^{i\delta\phi(t)}\rangle=\tilde{\rho}_{01}(0)e^{-\frac{1}{2}\langle\delta\phi^2(t)\rangle}\,,
\end{equation}
where $\langle\cdot\rangle$ denotes an ensemble average (over the random processes), and, for the general pulse sequence:
\begin{equation}
  \langle \delta\phi^2(t) \rangle = 4\pi^2 \int_{-\infty}^{+\infty} dt'\,\int_{-\infty}^{+\infty} dt''\,\sum_{i,j} D_{{\rm G}i} D_{{\rm G}_j}\langle \delta V_{{\rm G}i}(t') \delta V_{{\rm G}j}(t'')\rangle\filter(t')\filter(t'')\,.
\end{equation}
Under the assumptions that the noise on the different gates are independent, and that their respective auto-correlation functions are homogeneous in time, we reach in frequency domain:  
\begin{equation}\label{deltaphi}
  \langle \delta\phi^2(t)\rangle = 4\pi^2 \int_{-\infty}^{+\infty} df\, \sum_i D_{{\rm G}i}^2S_{{\rm G}i}(f)\left|\ffilter(f)\right|^2\,,
\end{equation}
where $S_{{\rm G}n}(f)=\int_{-\infty}^{+\infty} dt\,e^{-2i\pi ft}\left\langle \delta V_{{\rm G}n}(t) \delta V_{{\rm G}n}(0)\right\rangle$ is the Fourier transform of the auto-correlation function of the noise on gate G$n$ (the power spectrum according to the Wiener-Khinchin theorem), and $\ffilter(f)=\int_{-\infty}^{+\infty} dt\,e^{-2i\pi ft}\filter(t)$. Eq. (\ref{deltaphi}) can also be formalized using the filter function concept \cite{Biercuk2011, Cywinski_PRB_2008, Paladino_Revmodphys_2014}. We analyze below the different pulse sequences relevant for the present experiments.

\subsection{Free induction decay}

For the Ramsey sequence we have
\begin{equation}
\left|\ffilter^{\rm R}(f)\right|^2=\left(\frac{\sin(\pi ft)}{\pi f}\right)^2\,.
\end{equation}
Therefore, $\left|\ffilter^{\rm R}(f)\right|^2/t^2$ is close to unity up to $|f|\sim 1/t\sim1/T_2^*$, so that free induction decay is sensitive to noise in this whole range of frequencies. For low-frequency noise spectra of the form $S_{{\rm G}i}(f)=\Slf f_0/\max(|f|,f_l)$ together with a (soft) high-frequency cutoff $f_h$, we get in the regime $2\pi f_l\ll2\pi f_h\ll 1/t$:
\begin{equation}
     \exp\left(-\frac{1}{2}\langle\delta\phi_R(t)^2\rangle\right)\approx\exp\left[-4\pi^2 t^2 \ln\left(\frac{f_h}{f_l}\right)f_0\sum_i D_{{\rm G}i}^2\Slf\right] \equiv \exp\left[-\left(\frac{t}{T_2^*}\right)^2\right]\,, 
\end{equation}
with \cite{Ithier_PRB_2005}:
\begin{equation}\label{eqRamsey}
\frac{1}{T_2^*}\approx2\pi\sqrt{\ln\left(\frac{f_h}{f_l}\right)f_0\sum_i D_{{\rm G}i}^2\Slf}\,.
\end{equation}
As shown in the main text (Fig. 4), the averaged $T_2^*$ decreases with increasing $t_{\rm meas}\sim1/(2\pi f_{\rm l})$ as the experiment probes smaller and smaller noise frequencies.

We can also estimate the contribution of higher frequency noises with spectra $S_{{\rm G}i}(f)=\Shf\sqrt{f_0/f}$. The Ramsey oscillations then decay as
$
     \exp(-\frac{1}{2}\langle\delta\phi_R(t)^2\rangle)=\exp(-(t/T_{2,\rm hf}^*)^{3/2})
$,
where we define:
\begin{equation}
\frac{1}{T_{2,\rm hf}^*}=\left(\frac{16\pi^2}{3}f_0^{1/2}\sum_i D_{{\rm G}i}^2\Shf\right)^{2/3}\approx 14\left(f_0^{1/2}\sum_i D_{{\rm G}i}^2\Shf\right)^{2/3}\,.
\label{eq:T2starhf}
\end{equation}
The low-frequency and high-frequency contributions to the decay of the Ramsey signal cross over at time $t_*=T_2^*(T_2^*/T_{2,\rm hf}^*)^3\ll T_2^*$ when $T_2^*\ll T_{2,\rm hf}^*$, and the decay is dominated by the low-frequency noise when $t\gg t_*$.

\subsection{Hahn Echo sequence}

For the Hahn echo sequence,
\begin{equation}
\left|\ffilter^{\rm E}(f)\right|^2=\frac{\sin^4(\pi f t/2)}{(\pi f/2)^2}\,.
\end{equation}
Therefore, the integrand in Eq. (\ref{deltaphi}) is small at frequencies $|f|\ll 1/t$ and the integral is dominated by the region around $f_*=2/(\pi t)$ (with extent $\sim f_*$). $f_*$ is of the order of $10-100$\,kHz for Hahn-echo sequences with total length $t=10-100$\,$\mu$s. If in this range of frequencies the noise spectra are of the form $S_{{\rm G}i}(f)=\Shf(f_0/f)^\alpha$ ($0<\alpha\leq 2$ typically), then:
\begin{equation}
    \exp\left(-\frac{1}{2}\langle\delta\phi_E(t)^2\rangle\right)=\exp\left(-C_{\alpha}(2\pi t)^{\alpha+1}f_0^\alpha\sum_i D_{{\rm G}i}^2\Shf\right)\equiv\exp\left[-\left(\frac{t}{T_2^{\rm E}}\right)^{\alpha+1}\right]\,,
\end{equation}
where $C_\alpha=2\sin(\frac{\alpha\pi}{2})(2^{1-\alpha}-1)\Gamma(-1-\alpha)$, with $\Gamma$ the Gamma function \cite{NHMF}, and: 
\begin{equation}
 \frac{1}{T_2^{\rm E}}=2\pi\left(C_\alpha f_0^\alpha\sum_i D_{{\rm G}i}^2\Shf\right)^{\frac{1}{\alpha+1}}\,.
\end{equation}
In the particular case $\alpha=0.5$ (see main text), $C_{0.5}=\frac{4\sqrt{2\pi}}{3}(2^{1/2}-1)\approx 1.38$, so that:
\begin{equation}
    \frac{1}{T_2^{\rm E}}\approx7.8\left(f_0^{1/2}\sum_i D_{{\rm G}i}^2\Shf\right)^{2/3}\,.
\end{equation}
The Hahn echo $T_2^{\rm E}$ and Ramsey $T_{2,\rm hf}^*$ [Eq. (\ref{eq:T2starhf})] are thus proportional. Therefore, one would expect $T_{2,\rm hf}^*\simeq 50$\,$\mu$s at $\theta_{zx}=99\degree$ where $T_2^{\rm E}\simeq 90$\,$\mu$s if the limiting noise mechanisms were the same at low and high frequency. The much shorter $T_2^*$ measured in the present device hence support the existence of additional noise sources at low frequency. 

\subsection{CPMG sequence}

For the more general ${\rm CPMG}$ sequence \cite{Cywinski_PRB_2008,Medford_PRL_2012} with noise spectra $\Shf(f_0/f)^\alpha$ over extent $\sim 1/t$ around the frequency $f_{N_\pi}=N_\pi/(2t)\sim N_\pi/(2T_2^{\rm CPMG})$, we get the scaling
\begin{equation}
    \langle\delta\phi^2(t)\rangle\sim t^{\alpha+1}N_\pi^{-\alpha}f_0^{\alpha}\sum_iD_{{\rm G}i}^2\Shf\,,
\end{equation}
so that $\langle\delta\phi^2(t)\rangle\sim(t/T_2^{\rm CPMG})^{\alpha+1}$, with: 
\begin{equation}
    T_2^{\rm CPMG}\sim N_\pi^{\gamma}f_0^{-\gamma}\left(\sum_iD_{{\rm G}i}^{2}\Shf\right)^{-\frac{1}{\alpha+1}}
\end{equation}
and $\gamma=\alpha/(\alpha+1)$, in agreement with Ref. \citenum{Medford_PRL_2012}. 

\section{\label{suppinf:T2nonergodic}$T^*_2$ in the non ergodic regime}

In order to measure $T_2^*(\theta_{zx})$, we record $\approx 5.5$\,s long Ramsey oscillations over one hour for each magnetic field orientation. We vary the acquisition time by averaging $N$ consecutive traces ($t_{\rm meas}=N\times 5.5$\,s) and fit each of these data sets with a Gaussian decay where $T_2^*$ is a free parameter. Since the acquisition time can be faster than the low-frequency noise correlation time $\tau$ (non ergodic regime), $T_2^*$ is a stochastic variable that can be described by a statistical distribution \cite{Delbecq_PRL_2016}. 

\begin{figure*}[h!]
\includegraphics[width = 0.9\textwidth]{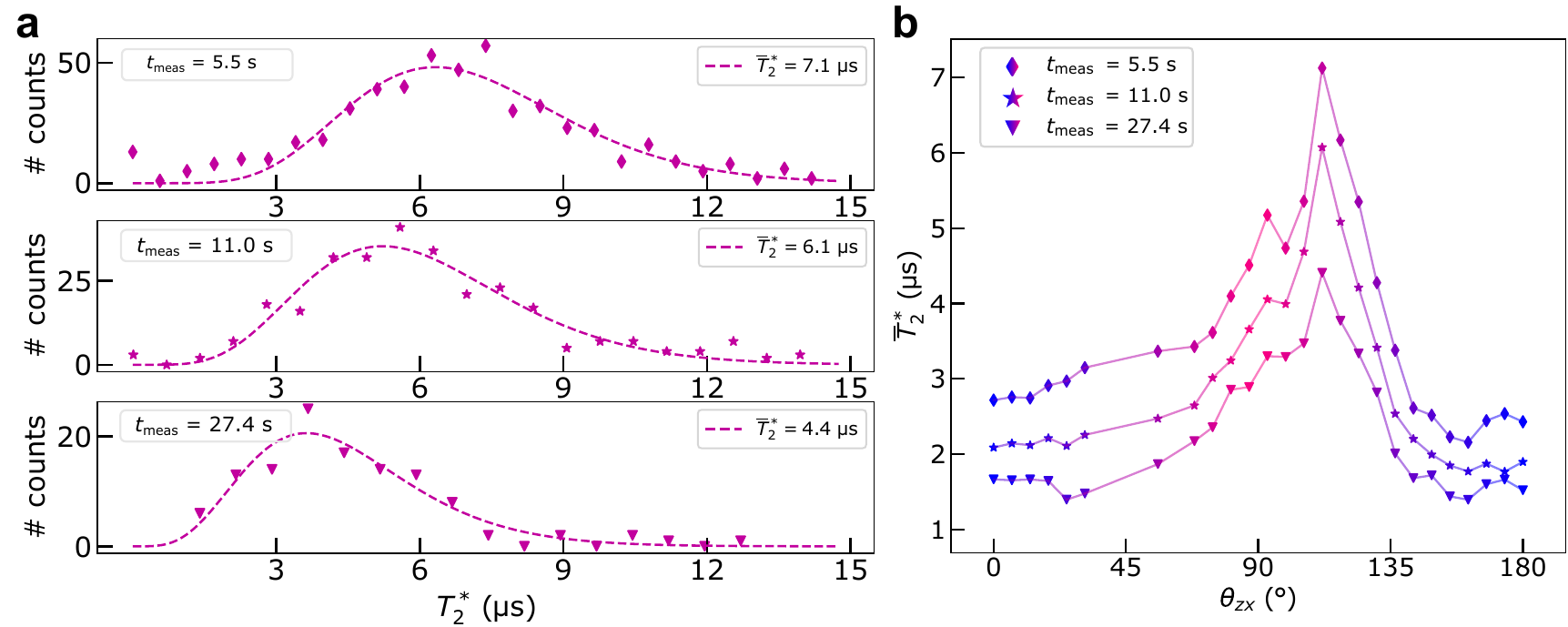}
\caption{\label{fig:t2star} 
\textbf{Spin coherence with correlated low-frequency noise} \textbf{(a)} $T_2^*$ histograms for $t_{\rm meas}=5.5$\,s ($N=1$, diamonds), $11$\,s ($N=2$, circles) and $27.4$\,s ($N=5$, squares) at $\theta_{zx}=111\degree$. The histograms are fitted with a Gamma distribution (dashed lines). \textbf{(b)} $\overline{T}_2^*$ extracted from the fits in (a) as a function of the magnetic field orientation $\theta_{zx}$.}
\end{figure*}

For $N \leq 5$, we can fit the histograms of $T_2^*$ with a Gamma distribution as in Ref. \citenum{Delbecq_PRL_2016} (see Fig \ref{fig:t2star}a):
\begin{equation}
f(T_2^*;\overline{T}_2^*,k)=\frac{k^k}{\overline{T}_2^{*^k}\Gamma(k)}T_2^{*^{k-1}}e^{-kT_2^*/\overline{T}_2^*}
\end{equation}
where $\Gamma$ is the Euler gamma function, $\overline{T}_2^*$ is the mean and $k$ describes the shape (skewness) of the distribution. 
The thus extracted $\overline{T}_2^*$ is more robust to the presence of $T_2^*$ data points far away from the mean, which are more frequent for small $N$'s. In Fig \ref{fig:t2star}b, we plot the fitted $\overline{T}_2^*$ as a function of the magnetic field orientation for $N=1,\,2,\,5$. The data exhibit a clear peak at $\theta_{zx}=111\degree$, close to (but not exactly at) the sweet spot of $T_2^{\rm E}$.

For $N>5$, the data set does not contain enough samples (less than 100) to extract the probability distribution parameters with high enough accuracy. In that case, we simply estimate $\overline{T}_2^*$ as the sample average of $T_2^*$. We point out that the distribution of $T_2^*$'s shall narrow when approaching the ergodic regime ($t_{\rm meas}\gg\tau$).

\section{\label{suppinf:hyperfine}Hyperfine interaction limit for the inhomogeneous dephasing time}

The hyperfine interactions between the hole and the $N$ nuclei spins are described by the following Hamiltonian \cite{Machnikowski2019,Bosco2021}:
\begin{equation}
H_{\rm int}=\frac{A}{2n_0}\sum_{n=1}^N\, \delta(\mathbf{r}-\mathbf{R}_n)\otimes\mathbf{J}\cdot\mathbf{I}_n\,,
\end{equation}
where $A$ is the hyperfine coupling constant, $n_0$ is the density of nuclei in the crystal, $\mathbf{I}_n$ is the spin operator of nuclei $n$ at position $\mathbf{R}_n$, and $\mathbf{J}$ is the angular momentum operator acting on the $J=3/2$ Bloch functions of the heavy and light holes (whereas the $\delta(\mathbf{r}-\mathbf{R}_n)$ acts on the envelopes). We discard here the small contributions from the split-off $J=1/2$ components as well as the small $\propto J_x^3,\,J_y^3,\,J_z^3$ corrections arising from the cubic symmetry of the  crystal \cite{Machnikowski2019}.

Let $\left|\uparrow\right\rangle$ and $\left|\downarrow\right\rangle$ be the pseudo-spin states of the dot at a given magnetic field, and $\left|\psi_{\rm nucl}\right\rangle$ be the nuclear configuration. The first-order correction to the Larmor energy $\varepsilon_L=hf_L$ is:
\begin{equation}
\delta\varepsilon_L=\frac{A}{2n_0}\sum_{n=1}^N\,\left\langle\psi_{\rm nucl}\right|\mathbf{I}_n\left|\psi_{\rm nucl}\right\rangle\cdot\big(\left\langle\uparrow\right| \delta(\mathbf{r}-\mathbf{R}_n)\otimes\mathbf{J}\left|\uparrow\right\rangle-\left\langle\downarrow\right| \delta(\mathbf{r}-\mathbf{R}_n)\otimes\mathbf{J}\left|\downarrow\right\rangle\big)\,.
\end{equation}
We next average over the nuclei configurations assuming uncorrelated and unpolarized nuclear spins with Gaussian-distributed quasi-static fluctuations \cite{Merkulov02}. The variance of $\delta\varepsilon_L$ is then:
\begin{equation}
\left\langle\delta\varepsilon_L^2\right\rangle=\frac{A^2}{4n_0^2}\sum_{n=1}^N\,\left\langle I_x^2\right\rangle\delta J_x^2(\mathbf{R}_n)+\left\langle I_y^2\right\rangle\delta J_y^2(\mathbf{R}_n)+\left\langle I_z^2\right\rangle\delta J_z^2(\mathbf{R}_n)\,,
\end{equation}
where, for $\alpha\in\{x,y,z\}$:
\begin{equation}
\delta J_\alpha(\mathbf{R}_n)=\left\langle\uparrow\right| \delta(\mathbf{r}-\mathbf{R}_n)\otimes J_\alpha\left|\uparrow\right\rangle-\left\langle\downarrow\right| \delta(\mathbf{r}-\mathbf{R}_n)\otimes J_\alpha\left|\downarrow\right\rangle\,,
\end{equation}
and $\left\langle I_x^2\right\rangle=\left\langle I_y^2\right\rangle=\left\langle I_z^2\right\rangle=I(I+1)/3$. Taking a second average over nuclei spin distributions, and assuming slowly varying envelope functions, we reach:
\begin{equation}
\left\langle\left\langle\delta\varepsilon_L^2\right\rangle\right\rangle=\frac{A^2}{12n_0}I(I+1)\nu\left(\overline{\delta J_x^2}+\overline{\delta J_y^2}+\overline{\delta J_z^2}\right)\,,
\end{equation}
where $\nu$ is the fraction of nuclei carrying a spin, and:
\begin{equation}
\overline{\delta J_\alpha^2}=\int d^3\mathbf{R}\,\delta J_\alpha^2\left(\mathbf{R}\right)\,.
\end{equation}
Finally, the rate of inhomogeneous dephasing due to hyperfine interactions is \cite{Fisher08,Testelin09}:
\begin{equation}
\Gamma_{2}^*=\frac{1}{T_{2}^*}=\frac{\sqrt{\left\langle\left\langle\delta\varepsilon_L^2\right\rangle\right\rangle}}{\sqrt{2}\hbar}=\frac{|A|}{2\hbar}\sqrt{\frac{\nu I(I+1)}{6n_0}}\left(\overline{\delta J_x^2}+\overline{\delta J_y^2}+\overline{\delta J_z^2}\right)^{1/2}\,.
\end{equation}
The above expression can be evaluated with the 6 bands $\mathbf{k}\cdot\mathbf{p}$ wave functions computed in section \ref{suppinf:Modeling}. For silicon, we use $n_0=49.94$ nm$^{-3}$, as well as $\nu=4.7\%$, $I=1/2$, and $|A|=1.67\,\mu$eV for $^{29}$Si isotopes \cite{Bosco2021}. This value of $|A|$ was specifically computed for holes with {\it ab initio} density functional theory \cite{Philippopoulos2020}. The resulting $T_{2}^*$, plotted as a dashed line in Fig. 4 of the main text, is minimal when the magnetic field $\mathbf{B}$ is along $y$, and maximal when it is in the $xz$ plane, as expected for a carrier that shows the strongest heavy-hole character when $\mathbf{J}$ is quantized along $y$. $T_{2}^*$ is weakly dependent on the angle $\theta_{zx}$, and is around $2.4\,\mu$s in the $xz$ plane.

\section{\label{suppinf:spectrum} Noise spectrum}

\begin{figure*}[h]
\includegraphics[width = 1 \textwidth]{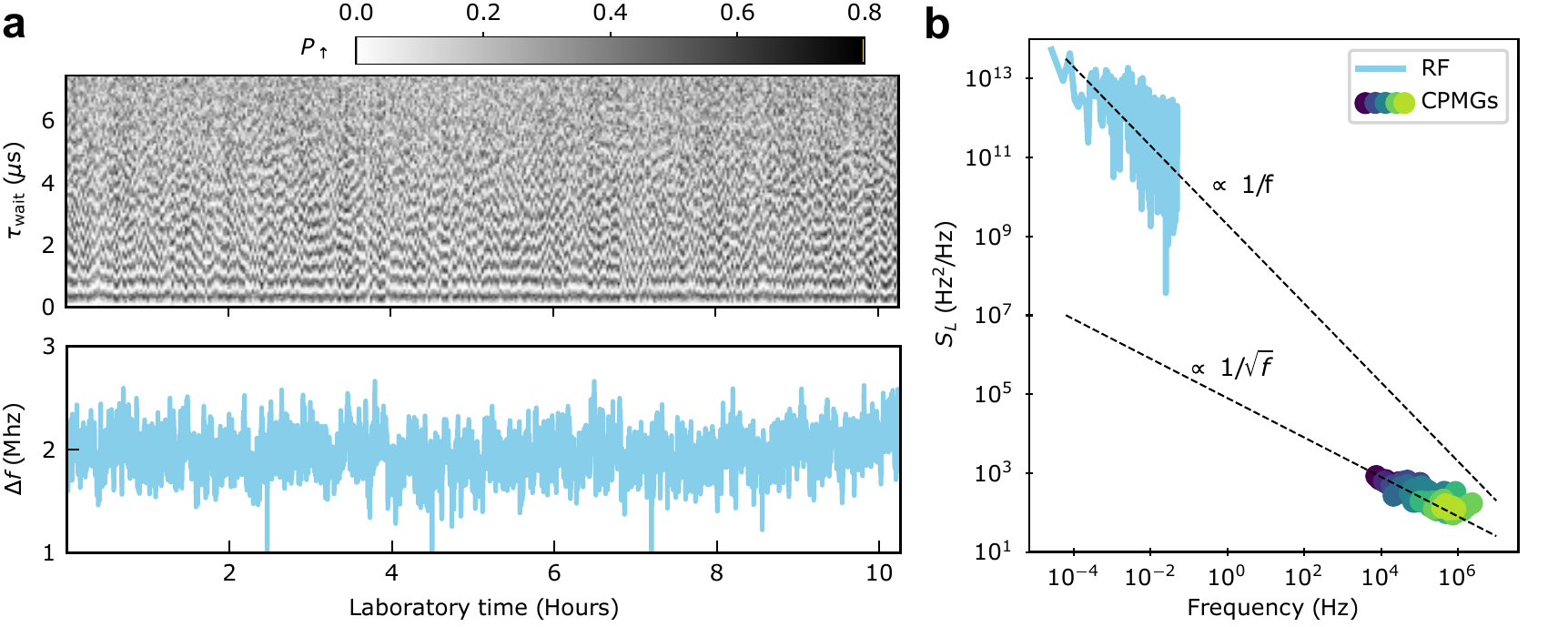}
\caption{\label{fig_spectrum} \textbf{Noise spectrum} \textbf{(a)} (top) Ramsey fringes as a function of $\tau_{\rm wait}$ acquired during 10 hours, at $\theta_{zx}=90\degree$. Each fringe oscillates at the frequency $\Delta f=f_{\rm MW1}-f_{\rm L}$. A single fringe takes roughly $10$\,s to record. (bottom) $\Delta f$, obtained via Fourier transform of the Ramsey fringes, versus laboratory time. \textbf{(b)} Power spectral density of the noise on the Larmor frequency. The low-frequency spectrum (RF) is calculated from (a) and is roughly proportional to $1/f$, as outlined by the upper dashed line. The high frequency spectrum (colored dots) is extracted from CPMG measurements with $N_\pi$ from 2 to 256, and is proportional to $1/f^{0.5}$ (lower dashed line).}
\end{figure*}

We measured 3700 Ramsey fringes over $t_{\rm tot}=10.26$ hours. For each realization, we varied the free evolution time $\tau_{\rm wait}$ up to $7\,\mu$s, and averaged 200 single shot spin measurement to obtain $P_{\uparrow}$ (see Fig.~\ref{fig_spectrum}a (top)). The fringes oscillate at the detuning $\Delta f=|f_{\rm MW1}-f_L|$ between the MW1 frequency $f_{\rm MW1}$ and the spin resonance frequency $f_L$. In order to track low-frequency noise on $f_L$, we make a Fourier transform of each fringe and extract its fundamental frequency $\Delta f$ reported in Fig.~\ref{fig_spectrum}a (bottom). During the whole experiment, $f_{\rm MW1}$ is set to $17$\,GHz.

The low-frequency spectral noise on the Larmor frequency (in units of $\rm Hz^2/Hz$) is calculated\footnote{Here we make use of two-sided power spectral densities, which are even with respect to the frequency.} from $\Delta f(t)$ as \cite{Yoneda-2018}:
\begin{equation}
S_L=\frac{t_{\rm tot}\left|{\rm FFT}[\Delta f]\right|^2}{N^2}\,,
\end{equation}
where ${\rm FFT}[\Delta f]$ is the fast Fourier transform (FFT) of $\Delta f(t)$ and $N$ is the number of sampling points. We observe that the low-frequency noise, plotted in Fig.~\ref{fig_spectrum}b, behaves approximately as $S_L(f)=S^{\rm lf}(f_0/f)$ with $S^{\rm lf}=10^9$\,Hz$^2$/Hz, which is comparable to what has been measured for a hole spin in natural Germanium \cite{Hendrickx_Nature_2020}. 

To further characterize the noise spectrum, we add the CPMG measurements as colored dots on Fig.~\ref{fig_spectrum}b \cite{Yoneda-2018}: 
\begin{equation}
S_L\left(N_\pi/(2 \tau_{\rm wait})\right)=-\frac{\ln(A_{\rm CPMG})}{2\pi^2\tau_{\rm wait}}\,,
\end{equation}
where $A_{\rm CPMG}$ is the normalized CPMG amplitude. As discussed in the main text, the resulting high frequency noise scales as $S^{\rm hf}(f_0/f)^{0.5}$, where $S^{\rm hf}=8\times 10^4$\,Hz$^2$/Hz is four orders of magnitude lower than $S^{\rm lf}$. This high frequency noise appears to be dominated by electrical fluctuations, as supported by the correlations between the Hahn-echo/CPMG $T_2$ and the LSESs. Additional quasi-static contributions thus emerge at low frequency, and may include hyperfine interactions (see section \ref{suppinf:hyperfine}).

\section{\label{suppinf:yield}Uniformity and quality of the samples at the wafer scale}

The devices were extensively characterized at room temperature prior to low-temperature measurements. 90\% of the 4-gate devices with 80 nm gate pitch (i.e. around 125 devices) are functional across the full 300 mm wafer. The devices are defined as functional according to 3 criteria:
\begin{itemize}
    \item With any gate G$i$ closed ($V_{{\rm G}i}=+0.2$ V) and the other gates G$j$ open ($V_{{\rm G}j}=-2$ V, $j\neq i$), the source-drain current $I_{\rm D}$ must be lower than $10^{-11}$ A at source-drain bias $V_{\rm DS}=50$ mV.
    \item With all gates open ($V_{{\rm G}i}=-2$ V), $I_{\rm D}$ must be greater than $10^{-7}$ A at $V_{\rm DS}=50$ mV.
    \item The gate leakage current $I_{{\rm G}i}$ must be lower than $10^{-11}$ A.
\end{itemize}

\begin{figure*}[t]
\includegraphics[width = 1\textwidth]{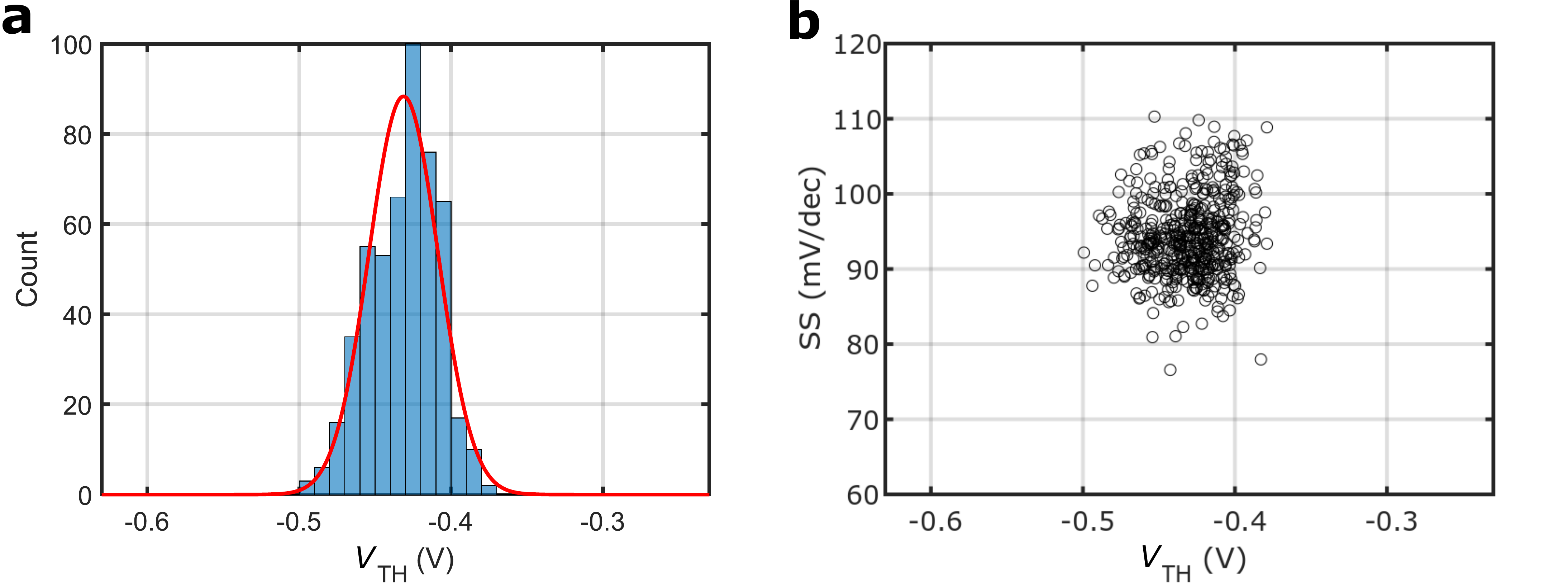}
\caption{
\textbf{Yield across the full wafer.} \textbf{(a)} Distribution of the room temperature threshold voltages $V_{\rm TH}$ of all gates of the devices similar to the one measured in this paper (four 40 nm long gates separated by 40 nm, on top of a 17 nm thick and 100 nm wide channel, with a 6 nm thick SiO$_2$ gate oxide). All 3 other gates are biased at $-2$ V. The red curve is a gaussian fit with average 0.43 V and  standard deviation 22 mV. $V_{\rm TH}$ is defined as the gate voltage where the derivative of the transconductance $\partial g_m/\partial V_{\rm G}$ is maximum. \textbf{(b)} Distribution of sub-threshold slope SS versus $V_{\rm TH}$ for all gates of all functional devices.}
\label{fig:yield}
\end{figure*}

Figure \ref{fig:yield}a collects the room temperature threshold voltages $V_{\rm TH}$ measured for each gate of each functional device (with $-2$ V applied on the 3 other gates). Figure \ref{fig:yield}b displays the sub-threshold slope (SS) versus the  threshold voltage $V_{\rm TH}$ of each gate.
The distribution of threshold voltages is sharply peaked around $V_{\rm TH}=-0.43$ V (standard deviation: 22 mV), which testifies the high uniformity of the devices at the wafer scale. As a comparison, the recent Ref. \citenum{Zwerver_2022} reports a standard deviation of up to 145 mV for the first gate layer. The uniformity of the devices in the wafer is further supported by the narrow distribution of sub-threshold slopes.

As compared to Ref. \citenum{Maurand_Ncomms_2016}, the fabrication process has been improved in several major aspects, that are described for instance in Ref. \citenum{Bedecarrats_IEDM_2021} (except for the exchange gates, that are not included in the present wafer):
\begin{itemize}
    \item The source and drain are now doped \textit{in situ} (during the overgrowth of the contacts). They were previously doped by ion implantation, which resulted in the spurious implantation of dopants in the channel.
    \item The source/drain junctions have been engineered to optimize the coupling with the reservoirs, including changes in the spacer design and thermal annealing step.
    \item High-k dielectrics (e.g. HfSiO$_2$) have been removed from the gate stack, leaving  SiO$_2$ as the only gate oxide. High-k dielectrics are known to host higher densities of charge traps, which can be very detrimental in the few-hole regime \cite{Biel_2021}.
    \item The silicon channel is thicker (17 nm) than in Ref. \citenum{Maurand_Ncomms_2016} (10 nm), which reduces the sensitivity to surface roughness \cite{Biel_2021}. 
\end{itemize}

\section{\label{suppinf:Setup} Setup}

\begin{figure*}[h!]
\includegraphics[width = 0.5 \textwidth]{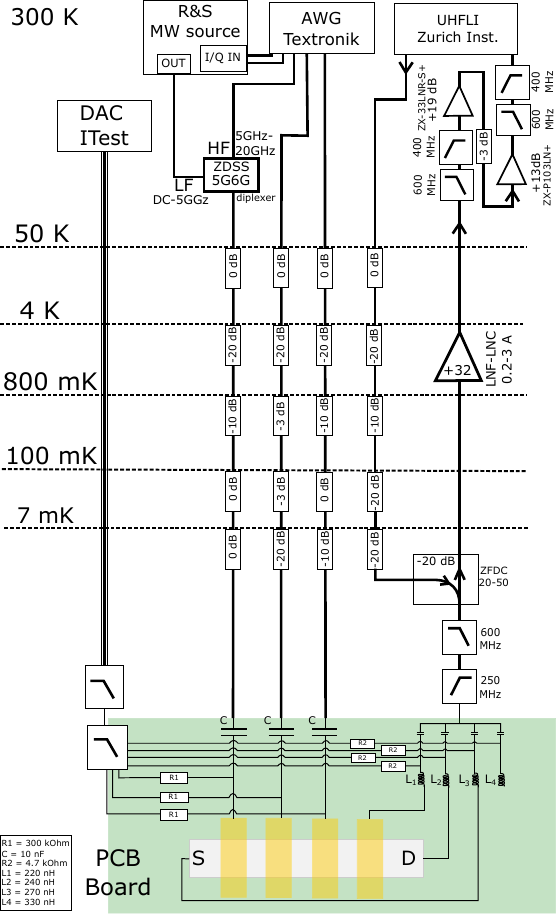}
\caption{\label{fig_setup} \textbf{Experimental setup.} Dilution fridge with all electrical connections to the sample.}
\end{figure*}

We operate in a dilution refrigerator system equipped with a three-axis vector superconducting magnet. The main solenoid magnet produces a magnetic field of up to 6\,T in the $z$ direction, while both transverse Helmholtz coils ramp up to 1\,T in the $x$ and $y$ directions. However, one of the axis was broken during the experiment. Therefore, after recording Fig. 1d of the main text, the sample was warmed up, physically rotated by $90\degree$, and cooled down again to record Fig. 1e. The electrical lines connecting the sample are displayed in Fig.~\ref{fig_setup}. 24 twisted pairs are filtered at the mixing chamber by 6 low pass filters. The DC gate voltages are generated by Itest high stability voltage sources (BE2141). To perform charge and spin manipulation, semi-rigid coaxial lines with $20$\,GHz bandwidth are routed to G1, G2 and G3 using on-PCB bias tees. Microwave frequency signals are supplied by a vector signal generator (R\&S SMW200A) with IQ modulating signals originating from two channels of an arbitrary waveform generator (AWG) Tektronix AWG5200. Other channels of the AWG are used to generate the pulse sequences. The homodyne readout of the resonator connected to the drain electrode is performed with a Zurich Instrument UHFLI lock-in with an excitation power of $-105$\,dBm at the PCB stage. The reflected signal from the resonator is amplified at $4$\,K with an ultra-low noise cryogenic amplifier LNF-LNC0.2-3A.

\end{document}